\documentclass[12pt]{article}
\usepackage{jheppub}

\usepackage{amsmath,amssymb,amsbsy}
\usepackage{graphicx,color}
\usepackage[dvipsnames]{xcolor}
\usepackage{cancel,slashed,mathrsfs}
\usepackage{hyperref}
\usepackage{framed}
\usepackage{cleveref}

\definecolor{shadecolor}{rgb}{0.90,0.90,0.90}

\graphicspath{{./figures/}}

\renewcommand{\d}{\mathrm{d}}
\newcommand{\be}{\begin{eqnarray}}
\newcommand{\ee}{\end{eqnarray}}
\newcommand{\bea}{\begin{eqnarray}}
\newcommand{\eea}{\end{eqnarray}}

\title{Energy-Energy Correlators at Strong Coupling}

\author[a]{Max Jackson,}
\author[a]{Lecheng Ren,}
\author[b,c]{Bo Wang,}
\author[a]{Congkao Wen}

\affiliation[a]{Centre for Theoretical Physics and Astronomy, Department of Physics and Astronomy,\\
  Queen Mary University of London, Mile End Road, London E1~4NS, United Kingdom}
\affiliation[b]{Zhejiang Institute of Modern Physics, School of Physics, Zhejiang University, \\
Hangzhou, Zhejiang 310058, China}
\affiliation[c]{Department of Physics, Yale University, New Haven, CT 06511}
\preprint{QMUL-PH-26-26}

\abstract{%
We study energy-energy correlators (EEC) in planar $\mathcal{N}=4$ super Yang-Mills theory at strong 't~Hooft coupling $\lambda$. We consider the EEC in states created by half-BPS operators of arbitrary dimension $p$, and determine the corresponding event-shape function up to order $\lambda^{-3/2}$ from the worldsheet representation of the AdS Virasoro-Shapiro amplitude with Kaluza-Klein external states. For $p=2$ we compute the second curvature correction, which completes the EEC through order $\lambda^{-2}$; the new contribution improves the agreement with recently derived non-perturbative bounds at intermediate coupling. We further develop a complementary method in which the strong-coupling expansion coefficients of the EEC are extracted directly from the Wilson coefficients of low-energy expansion of the AdS Virasoro-Shapiro amplitude, and find the two approaches in perfect agreement. 
}

\begin{document}
\maketitle

\section{Introduction}

Energy flow observables characterise the angular distribution of energy deposited at infinity in a scattering or decay process. First introduced in the late 1970s \cite{Sterman:1975xv,Basham:1978bw,Basham:1977iq}, energy correlators measure the normalised correlation between energy fractions collected by calorimeters 
on the celestial sphere, and play an important role in understanding the dynamics of collider experiments — from probing the structure of jets \cite{Basham:1978zq,Chen:2020vvp,Komiske:2022enw,Moult:2025nhu} and the onset of confinement in QCD \cite{Lee:2025okn,Guo:2025zwb,Chang:2025kgq,Kang:2025zto}, to precision tests of the Standard Model. In conformal field theories, where there is no intrinsic mass scale, energy correlators admit a precise non-perturbative formulation in terms of light-ray operators inserted at null infinity \cite{Hofman:2008ar,Kologlu:2019mfz}, and become amenable to powerful analytic techniques. This has provided an exciting and deep connection between collider physics, phenomenology, and formal QFT.

A particularly fruitful arena is $\mathcal{N}=4$ super Yang-Mills (SYM) theory. At weak coupling the energy-energy correlator (EEC) exhibits non-trivial power-law behaviour in the back-to-back and collinear limits, mirroring the jet structure familiar from QCD \cite{Belitsky:2013ofa,Belitsky:2013xxa,Belitsky:2013bja, Henn:2019gkr,Dixon:2019uzg}, while at strong coupling the EEC approaches the uniform angular distribution predicted holographically by supergravity scattering in AdS \cite{Hofman:2008ar}. Corrections to the uniform distribution are controlled by the AdS Virasoro-Shapiro amplitude -- the four-point amplitude of gravitons in $\mathrm{AdS}_5 \times S^5$, which has been bootstrapped in a small-curvature expansion \cite{Alday:2023jdk,Alday:2023mvu}, including external Kaluza-Klein modes \cite{Fardelli:2023fyq,Wang:2025pjo}. In \cite{Ren:2026zxs} a precise formula was established relating the worldsheet integral of the AdS Virasoro-Shapiro amplitude to the EEC at strong coupling; the coefficients of the AdS curvature expansion of the EEC are expressed as integrals of the worldsheet correlator over the unit disk and its boundary. This formula was used to compute the flat-space contribution and the first curvature correction, fixing the EEC through order $\lambda^{-3/2}$, for the correlator 
$\langle \mathcal{O}_2\,\mathcal{E}\,\mathcal{E}\,\mathcal{O}_2\rangle$, where $\mathcal{E}$ is the energy flux operator and $\mathcal{O}_2$ is the local half-BPS operator with protected conformal dimension two. Closely related strong-coupling results and non-perturbative bounds on the EEC have recently been obtained in \cite{Dempsey:2025yiv} from the conformal collider bootstrap. 

In this paper we extend the results of \cite{Ren:2026zxs} in several directions. Firstly, we consider energy correlations in states created by half-BPS operators $\mathcal{O}_p$ of arbitrary conformal dimension $p\geq 2$, characterised by the four-point function $\langle \mathcal{O}_p\,\mathcal{E}\,\mathcal{E}\,\mathcal{O}_p\rangle$. In general, the EEC is a function of the angle between two detectors $\theta$, and in the strong coupling expansion the EEC can be expressed as
\begin{equation} \label{eq:summary}
    \operatorname{EEC}_p(\xi) = 1 +\sum_{k=0}^{\infty} \lambda^{-\frac{k}{2}-1}  \sum_{n=0}^{k} c_{k,n} (p)  \, Q_{k-n,p}(\xi)   \,,
\end{equation}
where $\xi$ is related to the angle $\theta$ through the relation $\xi=\sin(\theta/2)^2$, and the angular functions $Q_{n,p}(\xi)$ are given in closed form in appendix~\ref{app: p22p Kernel Recurrence Relation}. We determine the coefficients $c_{k,n} (p)$ in the EEC by exploiting the precise connection between AdS Virasoro-Shapiro amplitude and the EEC established in~\cite{Ren:2026zxs}. The leading term $1$ in \eqref{eq:summary} is the supergravity contribution, and the higher order terms are the stringy corrections, which we will focus on. For generic $p$, utilising the results of~\cite{Fardelli:2023fyq} we determine first two orders of corrections, 
\begin{align}
c_{0,0}(p)&=\frac{\Gamma(p{+}3)}{\Gamma(p{-}1)\Gamma(p)}\,\zeta(2)\,,\qquad
c_{1,0}(p)=\frac{\Gamma(p{+}4)}{\Gamma(p{-}1)\Gamma(p)}\,\zeta(3)\,, \nonumber \\ 
c_{1,1}(p)&=\frac{\Gamma(p+3)}{\Gamma(p-1)\Gamma(p)} \bigl(\zeta(2) - (p^2+2p+2) \zeta(3)\bigr) \, . 
\end{align}
Secondly, for $p=2$ we further determine the $\mathcal{O}(\lambda^{-2})$ term of the EEC by applying the second curvature correction of the Virasoro-Shapiro amplitude~\cite{Alday:2023mvu}, with the relevant coefficients given as
\begin{align}
c_{2,0}  &= 855 \zeta(4) \, , \qquad c_{2,1}  = - 120\zeta(2) + 240 \zeta(3) - 3870\zeta(4) \, , \cr
c_{2,2} &= 304 \zeta(2) - 476 \zeta(3) + 24 \zeta(2)\zeta(3) + 4692 \zeta(4) - 40 \zeta(5)\, . 
\end{align}

We further develop a complementary method in which the coefficients of the EEC are obtained directly from the low-energy expansion of the Mellin amplitude, without recourse to the worldsheet formulation. This is achieved via two equivalent techniques: a Borel resummation of the Wilson coefficients, and a more efficient procedure based on analytic continuation and localisation of the summation variable to a specific value. The alternative method based on low-energy expansion reproduces the worldsheet results order by order in $1/\sqrt{\lambda}$, providing a non-trivial cross-check between the two independent approaches. Finally, we compare these analytic results against the numerical bounds from the conformal collider bootstrap~\cite{Dempsey:2025yiv}: at large values of the coupling the analytic results are consistent with the bootstrap bounds, and the second curvature correction computed here extends this agreement to smaller values of the coupling constant.

The rest of the paper is organised as follows: in section \ref{sec:EEC} we introduce the EEC in $\mathcal{N}=4$ SYM, emphasising its connection to the Mellin amplitudes of four-point correlators of half-BPS operators. In section \ref{sec:worldsheet}, we give a detailed relation between the energy-energy correlators and the worldsheet formulation of the AdS Virasoro-Shapiro amplitude, which provides a powerful method for deriving the strong coupling expansion of the EEC in $\mathcal{N}=4$ SYM. Explicit results are obtained for the EEC up to the order $\lambda^{-3/2}$ for operators with generic conformal dimension, and up to the order $\lambda^{-2}$ for operator of dimension two. In section \ref{sec:low-energy} we present an independent method of computing strong coupling expansion of the EEC using the low-energy expansion of the Mellin amplitudes. We find the results from these two approaches in perfect agreement. In section \ref{sec:strong-coupling}, we comment on physical properties of our results  and compare with the numerical bootstrap bounds. We conclude in section \ref{sec:con and out}, where we also comment on future directions and open questions. The paper also includes four appendices, which contain technical details for the results in the main text. Appendix \ref{app: p22p Kernel Recurrence Relation} provides a derivation for the angular functions $Q_{n,p}(\xi)$ appearing in the EEC, appendix \ref{app:worldsheet} reviews and solves the worldsheet integrals that arise at first and second curvature orders, appendix \ref{app:Wilson-coeff} provides details about the low-energy expansion coefficients of the Mellin amplitude, and finally in appendix \ref{app:ancont} we derive a set of localisation and regularisation rules to facilitate the derivation of the EEC from the low-energy coefficients.

\section{Energy-energy correlators in $\mathcal{N}=4$ SYM: four-point functions and the detector kernels}
\label{sec:EEC}

The energy flow operator measures the total energy flux through null infinity in the direction of the unit vector $n= (1, \hat{n})$,
\begin{equation}
\label{eq:Edef}
    \mathcal{E}(n) = \lim_{r \to \infty} r^2 \int_{0}^{\infty} \d t \, n^i \, T_{0i}(t, r \hat{n}),
\end{equation}
where $T_{\mu\nu}$ is the stress tensor of the theory and $n$ is null. The energy-energy correlator in a state created by the local source operator $\mathcal{O}$ is defined as the event-shape observable
\begin{equation}
\label{eq:EECdef}
    \operatorname{EEC}(\xi) = \frac{16\pi^2}{q^2\,\sigma_{\rm tot}(q)}\int \d^4x\; e^{iq\cdot x}\,\langle \mathcal{O}(x)\,\mathcal{E}(n_1)\,\mathcal{E}(n_2)\,\mathcal{O}(0)\rangle,
\end{equation}
where the correlator is a Wightman function, and $\sigma_{\rm tot}(q)=\int \d^4x\,e^{iq\cdot x}\langle \mathcal{O}(x)\mathcal{O}(0)\rangle$ is the total cross-section for the source to create any final state of total momentum $q$. The EEC depends only on the angular variable 
\begin{equation}
    \xi = \frac{q^2\,(n_1 {\cdot} n_2)}{2\,(q {\cdot} n_1)(q {\cdot} n_2)}\, , 
\end{equation}
which, in the source rest frame, becomes $\xi = \frac{1}{2}(1-\cos\theta)$ with $\theta$ denoting the angle between the detectors.

We consider the  EEC in  $\mathcal{N}=4$ SYM with gauge group $SU(N)$ in the planar limit. We take the source operators in \eqref{eq:EECdef} to be the single-trace half-BPS superconformal primaries, which in terms of fundamental fields are defined as
\begin{equation}
    \mathcal{O}_p(x,Y) = \operatorname{tr}\big[\phi^p(x)\big], \qquad \phi(x)=\sum_{I=1}^6 Y^I \Phi^I(x),
\end{equation}
where $\Phi^I$, $I=1,\ldots,6$, denote the six real scalar fields of the theory and $Y$ is a null $SO(6)_R$ polarisation vector. The operator $\mathcal{O}_p$ has protected scaling dimension $p$ and transforms in the $[0,p,0]$ representation of $SO(6)_R \cong SU(4)_R$. In the dual string theory these operators correspond to Kaluza--Klein modes on $\mathrm{AdS}_5 \times S^5$.

Importantly, the stress tensor belongs to the same supermultiplet as $\mathcal{O}_2$, and therefore it is possible to use superconformal Ward identities~\cite{Belitsky:2014zha} to relate the four-point function $\langle \mathcal{O}_p \, \mathcal{E} \, \mathcal{E} \, \mathcal{O}_p\rangle$ to the scalar correlator $\langle \mathcal{O}_p \, \mathcal{O}_2 \, \mathcal{O}_2 \, \mathcal{O}_p\rangle$. We begin by considering the scalar correlator in the Euclidean space, which can be written in the following factorised form~\cite{Eden:2000bk, Nirschl:2004pa}
\begin{equation}
\label{eq:p22p}
 \langle \mathcal{O}_p(1)\,\mathcal{O}_2(2)\,\mathcal{O}_2(3)\,\mathcal{O}_p(4)\rangle_E =  G^{\rm Born}_{p22p}(x_i, Y_i) + \mathcal I_{1234} \, d_{14}^{p-2} \, \frac{\Phi_p(u, v)}{x_{14}^2 x_{23}^2},
\end{equation}
where $\mathcal{O}_p(1) \equiv \mathcal{O}_p(x_1, Y_1)$ and the Born contribution is determined from the free theory. Here, $\mathcal I_{1234}$ is a universal rational prefactor, which will not be important for our discussion, and we have defined $d_{ij} = Y_i {\cdot} Y_j/x_{ij}^2$ to be the free scalar propagator with $x_{ij} \equiv x_i - x_j$. All dynamical information is now contained within the reduced correlator, $\Phi_p(u, v)$, a conformally invariant function of the coupling and the cross-ratios
\begin{equation}
    u = \frac{x^2_{12}x^2_{34}}{x^2_{14}x^2_{23}}, \qquad v = \frac{x^2_{13}x^2_{24}}{x^2_{14}x^2_{23}}\, .
\end{equation}
The function $\Phi_p(u,v)$ admits the following Mellin representation~\cite{Mack:2009mi, Penedones:2010ue}
\begin{equation}
\label{eq:MellinPhi}
    \Phi_p(u,v) = \frac{1}{4p}\int_{-i \infty}^{i \infty} \frac{\d s\,\d t}{(2\pi i)^2}\, u^{\frac12(\tilde u - p)}\,v^{\frac12(s-p)}
    \Gamma^2\Big(\tfrac{p+2-s}{2}\Big)\Gamma\Big(2-\tfrac{t}{2}\Big)\Gamma\Big(p-\tfrac{t}{2}\Big)\Gamma^2\Big(\tfrac{p+2-\tilde u}{2}\Big)\,\mathcal{M}_p(s,t)\, ,
\end{equation}
with $s+t+\tilde u = 2p$, and where both integration contours lie entirely to the left of the poles in the Gamma functions. We further introduce the Borel transform of the Mellin amplitudes~\cite{Penedones:2010ue}
\begin{align}
    \mathcal{M}_p(\hat{s}, \hat{t} ) = \frac{1}{2 \lambda^{3/2} \Gamma(p{-}1)\Gamma(p)}  \int_0^{\infty} \d\beta \, \beta^{p+3} \, e^{-\beta} A(\beta \hat{s}, \beta \hat{t}) \, , 
\end{align}
and we have expressed the amplitudes in terms of new variables, 
\begin{equation} \label{eq:that}
    \hat{s} = \frac{s_1}{2\sqrt{\lambda}} = \frac{s-2p/3}{2\sqrt{\lambda}}\, , \qquad \hat{t} = \frac{s_2}{2\sqrt{\lambda}} = \frac{t-2p/3}{2\sqrt{\lambda}}\, ,
\end{equation}
and $s_3=\tilde u-2p/3$ with $s_1+s_2+s_3=0$. Importantly, $A(S, T)$ is precisely the generalisation of the Virasoro-Shapiro amplitude in AdS space, with the flat-space limit given by \cite{Virasoro:1969me, Shapiro:1970gy}
\begin{equation}
\label{eq: Flat space VS}
    A^{(0)}(S, T) = -\frac{\Gamma(-S) \Gamma(-T) \Gamma(S+T)}{\Gamma(S+1) \Gamma(T+1) \Gamma(1-S-T)} \, . 
\end{equation}
The Virasoro-Shapiro amplitude will play an important role in the study of the EEC.

Recall from above that in $\mathcal{N}=4$ SYM the EEC can be directly related to the four-point correlator $\langle \mathcal{O}_p \mathcal{O}_2 \mathcal{O}_2 \mathcal{O}_p\rangle$. The Ward identities relate the EEC to the scalar flow correlator (SSC) in the $\mathbf{105}$ R-symmetry channel by setting $Y_2 = Y_3$: the latter fixes the scalar function, and the EEC follows by applying the spin-two Ward-identity at the detector points \cite{Chicherin:2023gxt}. This relation is most conveniently expressed using Mellin amplitudes~\cite{Belitsky:2013xxa, Belitsky:2013bja, Chicherin:2023gxt}. In particular the Mellin representation provides a natural pathway for understanding the Wick rotation from the Euclidean correlator to the Wightman function. Explicitly,  the EEC can be written in the following form
\begin{equation}
\label{eq:EECp def}
     \operatorname{EEC}_p(\xi) = \int^{i \infty}_{-i \infty} \frac{\d s\, \d t}{(2\pi i)^2} \mathcal{M}_p(s, t) \, \mathcal{K}_p(t; \xi)\, ,
\end{equation}
where the detector kernel $\mathcal{K}_p(t; \xi)$ results from acting with the $\mathcal{N}=4$ Ward identities, Wick rotating to the Wightman correlator and then performing the Fourier integral. It is given by the following sum \cite{Chicherin:2023gxt}
\begin{multline}
\label{eq:kernel for general spins}
    \mathcal{K}_p(t;\xi) = \sum_{k=0}^{2} {2 \choose k} \frac{(-1)^k \xi^{\frac{t}{2}-k} \pi^2 t (t-2) \Gamma(k{-}\frac{t}{2}) \Gamma(p{-}1) \Gamma(p) \Gamma(p{-}\frac{t}{2})}{32 \sin^2(\frac{\pi t}{2}) \Gamma(p{+}k{-}2{-}\frac{t}{2}) \Gamma(k{-}2{-}\frac{t}{2}) \Gamma(-2{-}\frac{t}{2}) \Gamma(p{-}k{+}3{+}\frac{t}{2})} \\ \times \, _2F_1\Bigl(3-k+\frac{t}{2}, 3-k+\frac{t}{2}, p-k+3+\frac{t}{2}; \xi\Bigr)\, ,
\end{multline}
where we have normalised the EEC such that the leading (supergravity) term of the strong coupling expansion is one. In the special case $p=2$ this can be used to write the detector kernel in a particularly simple form~\cite{Belitsky:2013bja}
\begin{equation}
    \mathcal{K}_2(t;\xi) = -\frac{\pi t^2 (t-2)^2}{128 \xi^3 \sin(\pi t/2)}\left(\frac{\xi}{1-\xi}\right)^{1+\frac{t}{2}}.
\end{equation}

In the strong coupling expansion, the Mellin amplitude $\mathcal{M}_p(s,t)$ is given by a leading supergravity contribution followed by an infinite tower of stringy corrections
\begin{equation}
\label{eq: Mellin amp expansion}
    \mathcal{M}_p(s, t) = \frac{4p}{\Gamma(p{-}1)} \frac{1}{(s-p)(t-2) (p-s-t)} + \frac{\Gamma(4{+}p)}{\Gamma(p) \Gamma(p{-}1)} \frac{\zeta(3)}{\lambda^{3/2}} + \ldots\, . 
\end{equation}
Although the physical EEC is expected to be finite, the Mellin integrals \eqref{eq:EECp def} associated with individual terms in this expansion are separately not well-defined (except the supergravity contribution). More precisely, a naive term-by-term evaluation can lead to divergences in the $s$-integration. This reflects the fact that the strong-coupling expansion of the correlators and light-ray transform \eqref{eq:Edef} do not commute. A well-defined strong-coupling expansion of the EEC therefore requires an alternative organising principle of the Mellin amplitudes rather than a simple strong coupling expansion as in \eqref{eq: Mellin amp expansion}. In the next section we show that such a principle is furnished by the worldsheet formulation of the AdS Virasoro-Shapiro amplitude.

\section{Energy-energy correlators from worldsheet integrals} \label{sec:worldsheet}

We now turn to the worldsheet formulation of the AdS Virasoro-Shapiro amplitude, which expresses each order of the curvature expansion as a well-defined integral over the moduli space of the string worldsheet. Relating the EEC to the worldsheet formulation allows the $s$-integration to be carried out before any expansion in $1/\sqrt\lambda$ is performed, and thereby provides a systematic and unambiguous method for determining the strong coupling coefficients of the EEC.

\subsection{Method outline}
\label{sec:method}

Exploiting the relationship between the Mellin amplitude and the EEC in $\mathcal{N}=4$ SYM (\ref{eq:EECp def}) and further use the Borel transform \eqref{eq:MellinPhi}, we obtain
\begin{equation}
\label{eq: EEC as Borel transform 1}
    \operatorname{EEC}_p\left(\xi\right) = \frac{1}{2 \lambda^{3/2} \Gamma(p{-}1)\Gamma(p)} \int_{-i \infty}^{i \infty} \frac{\d s \, \d t}{(2\pi i)^2} \int_0^{\infty} \d\beta \, \frac{\beta^{p+3}}{e^\beta} A(\beta \hat{s}, \beta \hat{t}) \, \mathcal{K}_p(t;\xi)\, ,
\end{equation}
which expresses the EEC in terms of AdS Virasoro-Shapiro amplitude $A(S, T)$. The Virasoro-Shapiro amplitude admits a large-curvature expansion
\begin{align} \label{eq:A-expansion}
    A(S, T) = \sum_{k=0}^\infty\lambda^{-k/2} A^{(k)}(S, T) \, , 
\end{align}
where $S = \beta \hat{s}$, $T = \beta \hat{t}$, (with $\hat s$ and $\hat t$ as defined in \eqref{eq:that}), and we have expressed the expansion in terms of  the 't Hooft coupling $\lambda$, which is related to the AdS curvature radius $R$ as $1/\sqrt{\lambda} = \alpha'/R^2$ by the AdS/CFT dictionary.

Unlike the $s$ integral we commented on earlier, due to the detector kernel $\mathcal{K}_p(t; \xi)$  the $t$ integral can be done by expanding $A(\beta \hat{s}, \beta \hat{t})$ in small $t$ without encountering any subtleties. This allows us to organise the EEC in the strong coupling regime as follows~\cite{Goncalves:2014ffa, Ren:2026zxs}
\begin{equation}
\label{eq: EEC_strong_from_Q}
    \operatorname{EEC}_p(\xi) = 1 + \sum_{k=0}^\infty \lambda^{-\frac k2 -1} \sum_{n=0}^k c_{k,n}(p) \, Q_{k-n, p}(\xi)\, ,
\end{equation}
where the dependence on the angular variable $\xi$ is entirely contained within the functions
\begin{equation}
    Q_{n,p}(\xi) = \int_{-i \infty}^{i \infty} \frac{\d t}{2\pi i} t^n \mathcal{K}_p(t; \xi)\, .
\end{equation}
For $p=2$ it is easy to check that $Q_{0,2}(\xi) = 1-6\xi +6\xi^2$, and furthermore that
\begin{equation}
    Q_{n,2}(\xi) = \frac{\partial}{\partial\xi} \bigl(2\xi(1-\xi) Q_{n-1,2}(\xi)\bigr) + 2(1-\xi) Q_{n-1,2}(\xi)\, .
\end{equation}
This recurrence relation can be generalised to all $p \geq 2$. In Appendix~\ref{app: p22p Kernel Recurrence Relation} we define and solve this relation to give a closed-form expression for $Q_{n,p}(\xi)$, given in \eqref{eq:Qnp}.
For example, terms of order $\mathcal{O}(\hat{t}^2)$ in $A(\beta \hat{s}, \beta \hat{t})$ become quadratic polynomials in $t$, and hence after integrating over $t$ we will obtain three contributions from $Q_{0,p}(\xi)$, $Q_{1,p}(\xi)$ and $Q_{2,p}(\xi)$ respectively.

What we require now is the AdS Virasoro-Shapiro amplitude $ A(S, T)$. The leading contribution is the flat-space Virasoro-Shapiro amplitude as quoted in \eqref{eq: Flat space VS}. For the higher order curvature corrections we make use of a remarkable result proposed in~\cite{Alday:2023jdk, Alday:2023mvu}, wherein it was argued that each term in the large-curvature expansion can be written as an integral over the Riemann sphere
\begin{equation}
\label{eq: AdS VS Worldsheet Integral}
    A^{(k)}(S, T) = \int \d^2z \, |z|^{-2S-2} \, |1-z|^{-2T-2} \, G^{(k)}_{\text{tot}} (z, \bar z; S, T)\, ,
\end{equation}
where $\d^2z = \frac{\d z \wedge \d\bar{z}}{-2\pi i}$ and the worldsheet correlator $G^{(k)}_{\text{tot}} (z, \bar z; S, T)$ is a single-valued multiple polylogarithm (SVMPL) of transcendental weight $3k$ \cite{Alday:2023mvu}, the details of which can be found in Appendix~\ref{app: SVMPLs}. The worldsheet correlator has been bootstrapped up to second curvature correction in the case $p=2$~\cite{Alday:2023mvu}, and up to first curvature correction for sources of general dimension $p >2$~\cite{Fardelli:2023fyq}. 

By exchanging the Borel integral with the Mellin-Barnes integral in (\ref{eq: EEC as Borel transform 1}), we can neatly express the EEC in the following form
\begin{equation}
\label{eq: EEC_in_terms_of_I}
    \operatorname{EEC}_p(\xi) = \frac{1}{\lambda} \frac{1}{\Gamma(p{-}1)\Gamma(p)} \int_{-i \infty}^{i \infty} \frac{\d t}{2\pi i} \int_0^\infty \d\beta \, e^{-\beta} \beta^{p{+}2} I(\beta\hat{t}) \, \mathcal{K}_p(t;\xi)\, ,
\end{equation}
where we have defined 
\begin{equation}
\label{eq: I(T) def}
    I(T) = \int_{-i \infty}^{+i \infty} \frac{\d s}{2\pi i} \frac{\beta}{2\sqrt{\lambda}} A(\beta \hat{s}, \beta \hat{t}) = \int_{-i \infty}^{+i \infty} \frac{\d S}{2\pi i}A(S, T) \, , 
\end{equation}
and from the strong coupling expansion of $A(S,T)$ given in \eqref{eq:A-expansion} we have the large-$\lambda$ expansion of $I(T) = \sum_{k=0}^{\infty} \lambda^{-k/2} I^{(k)}(T)$, with
\begin{equation}
    I^{(k)}(T) = 
    \int_{-i \infty}^{+i \infty} \frac{\d S}{2\pi i} A^{(k)}(S,T)\, .
\end{equation}
The above relation shows that $I^{(k)}(T)$ fully determines the EEC up to $\mathcal{O}(\lambda^{-1-\frac k2})$. In particular the strong coupling coefficients $c_{k,n}$ from \eqref{eq: EEC_strong_from_Q} are given by the small $T= \beta \hat t$ expansion of $I(T)$. Note that we should drop all of the singular terms in $T$ because they are related to the supergravity contribution, which is given as $1$ in \eqref{eq: EEC_strong_from_Q} and we will omit its contribution in the following discussions. Using the definition \eqref{eq:that}, we can write $I^{(k)}(T)$ as an expansion in $t$:   
\begin{equation}\label{eq:expandI}
    I^{(k)}(T) = \sum_{i\geq j \geq 0}^{\infty} d_{i,j}^{(k)} \, \left(\frac{\beta}{\sqrt{\lambda}}\right)^i t^j\, .
\end{equation}
Substituting this expression into \eqref{eq: EEC_in_terms_of_I}, the integral over $\beta$ will then map $\beta^i \to \Gamma(p+3+i)$ and $t^j \to Q_{j,p}(\xi)$. From the identity
\begin{align}
\sum_{k = 0}^{\infty} \lambda ^{- \frac{k}{2}} \sum_{n = 0}^{k} c_{k , n} \, Q _{k-n, p} (\xi)= \sum_{m = 0}^{\infty} \lambda ^{- \frac{m}{2}} \sum_{i \geq j \geq 0}^{\infty} \lambda ^{- \frac{i}{2}}  \frac{\Gamma(p+3+i) }{\Gamma(p{-}1)\Gamma(p)} d^{(m)}_{i , j} \, Q_{j, p}(\xi) \, ,
\end{align}
and by identifying $j=k-n$, we conclude that 
\begin{equation}\label{eq:relationcandd}
    c_{k,n} = \sum_{i=k-n}^{k} \frac{\Gamma(p+3+i) }{\Gamma(p{-}1)\Gamma(p)} \, d^{(k-i)}_{i,k-n}\,. 
\end{equation}

Given the form of the worldsheet ansatz for $A^{(k)}(S,T)$ in (\ref{eq: AdS VS Worldsheet Integral}), we see that $I^{(k)}(T)$ may be expressed as a sum of integrals of the form
\begin{equation}
    \mathcal{J}_j(T) = \int_{-i \infty}^{i \infty} \frac{\d S}{2\pi i} \int \d^2z \, S^j |z|^{-2S-2} J(z, \bar{z}; T)\, ,
\end{equation}
where $J(z, \bar z; T)$ is the remaining parts of the worldsheet correlator after stripping the $S$ dependence. We can sort integrals of this type into two cases, depending on whether the exponent $j$ is smaller than zero. 

\paragraph{Case 1: $j < 0$.}
Using the Schwinger parameterisation
\begin{equation}
\label{eq:Integral case 1}
    S^{j} = \frac{1}{\Gamma(-j)} \int_0^\infty \d\alpha \, \alpha^{-j-1} e^{-\alpha S} \, ,
\end{equation}
and writing $|z|^{-2S} = e^{-S\log |z|^2}$, the $S$ integral yields the delta function constraint $\delta(\alpha + \log |z|^2)$. The resulting moduli space integral is thus restricted to the unit disk $|z|<1$. Therefore,
\begin{equation}\label{eq:bulkgeneral}
    \mathcal{J}_{j}(T) = \frac{1}{\Gamma(-j)} \int_{|z|<1} \d^2z \,  (-\ln |z|^2)^{-j-1} |z|^{-2} J(z, \bar{z}; T)\,,\quad j <0 \,.
\end{equation}

\paragraph{Case 2: $j \geq 0$.}
The $S$ integral in this case is even simpler
\begin{equation}
    \int \frac{\d S}{2\pi i} S^j e^{-S\log |z|^2} = (-1)^j \, \delta^{(j)}(\log |z|^2)\, ,
\end{equation}
and we see that the integral localises on the unit circle. Changing to polar coordinates $z=re^{i\theta}$, we obtain
\begin{equation}\label{eq:boundarygeneral}
    \mathcal{J}_j(T) = \oint_{|z|=1} \frac{\d\theta}{2\pi} \left(\frac r2 \frac{\partial}{\partial r} \right)^j J(z, \bar{z}; T)\,, \quad  j \geq 0 \,.
\end{equation}

Having set out the general prescription, we now apply it to obtain explicit results for the strong-coupling expansion of the EEC. 

\subsection{Flat-space contributions}

We begin by considering the simplest case: the flat space contributions. We can integrate (\ref{eq: Flat space VS}) directly and obtain
\begin{equation}
    \label{eq: flat space worldsheet integral}
    I^{(0)}(T) = \int_{-c-i \infty}^{-c+i \infty} \frac{\d S}{2\pi i}A^{(0)}(S, T) = \frac{1}{T^2} \left(\frac{\Gamma(1-2T)}{\Gamma(1-T)^2} -1\right) \, ,
\end{equation}
where $c$ is a small positive number to specify the contour. Since $T \propto 1/\sqrt{\lambda}$, an expansion of $I^{(0)}(T)$ in small $T = \beta \hat t$ gives the strong-coupling expansion of the EEC.  Explicitly, we have, 
\begin{align}
\label{eq: flat space worldsheet integral small T}
    I^{(0)}(T) &= \zeta(2) + 2\zeta(3) T + \frac{19}{4} \zeta(4) T^2 + \mathcal{O}(T^3) \cr &= \zeta(2) + \frac{\beta}{\sqrt{\lambda}} \left(t-\frac{2p}{3}\right) \zeta(3) + \frac{19}{16} \frac{\beta^2}{\lambda} \left(t-\frac{2p}{3}\right)^2 \zeta(4) + \cdots \, ,
\end{align}
where we have used $T = \beta \hat{t}$ and the relation \eqref{eq:that}. 

The flat-space result determines the leading stringy correction to the EEC, which contributes at order $\mathcal{O}(\lambda^{-1})$. After applying the $\beta$ integral map and the rule (\ref{eq:relationcandd}), it is then easy to see that
\begin{align}
\label{eq: c00 p22p}
    c_{0,0} = \frac{\Gamma(p+3)}{\Gamma(p{-}1) \Gamma(p)} \zeta(2)\, ,
\end{align}
and the EEC at $\mathcal{O}(\lambda^{-1})$ is given by
\begin{align}
    \label{eq:leading corr}\operatorname{EEC}_p(\xi) \Big|_{\mathcal{O}(\lambda^{-1})} = c_{0,0} \, Q_{0,p}(\xi) = 24\zeta(2)  \, Q_{0,2}(\xi)\, ,
\end{align}
where we have used \eqref{eq:Q0pxi}, and $Q_{0,2}(\xi) = 6\xi^2-6\xi+1$. Interestingly, the leading stringy correction is $p$-independent, in agreement with~\cite{Chicherin:2023gxt}. In fact, all the coefficients $c_{k,0}$ with $k \geq 0$ in \eqref{eq: EEC_strong_from_Q} are completely fixed by the expansion (\ref{eq: flat space worldsheet integral small T}) via (\ref{eq:relationcandd}). Below are the first few examples beyond $c_{0,0}$: 
\begin{equation}
\label{eq: Flat space coeffs p22p from worldsheet}
 c_{1,0} = \frac{\Gamma(p+4)}{\Gamma(p{-}1) \Gamma(p)} \zeta(3)\,, \quad c_{2,0} = \frac{19\, \Gamma(p+5)}{16 \, \Gamma(p{-}1) \Gamma(p)} \zeta(4)\,, \,  \ldots
\end{equation}

The coefficients $c_{k,n}$ with $n>0$ receive contributions from flat space but also from the higher-curvature corrections to the Virasoro-Shapiro amplitude. For example, both the flat space and first curvature correction contribute to $c_{1,1}$. This is made manifest by considering the form of \eqref{eq:relationcandd}, where we identify the $i$-th curvature contribution to the coefficient $c_{k,n}$ to be
\begin{equation}\label{eq:cknseparate}
    c_{k,n}^{(i)} = \frac{\Gamma(p+3+k-i) }{\Gamma(p{-}1)\Gamma(p)} \, d^{(i)}_{k-i,k-n}\,. 
\end{equation}
In this way we can identify the flat-space contribution to $c_{1,1}$ from \eqref{eq: flat space worldsheet integral small T}, which is
\begin{equation}
\label{eq: c11 from flat space worldsheet}
    c_{1,1}^{(0)} = -\frac{2p \, \Gamma(p+4) \zeta(3)}{3 \Gamma(p-1) \Gamma(p)}\, .
\end{equation}
To completely determine $c_{1,1}$ (i.e.  $c_{1,1}^{(0)} + c_{1,1}^{(1)}$) requires the first-curvature correction to the Virasoro-Shapiro amplitude, which we will discuss in the next subsection. 

It is intriguing to note that $I^{(0)}(T)$ in \eqref{eq: flat space worldsheet integral} takes a similar form to graviton scattering off D-branes \cite[eq.~(29)]{Hashimoto:1996bf}. Moreover, interestingly the structure $\zeta(2)/\lambda$ in the leading stringy correction \eqref{eq:leading corr} also arises in certain correlator in $\mathcal{N}=4$ SYM \cite{Brown:2024tru, Brown:2026dhy}: more precisely the giant graviton correlator, which holographically describes the gravitons scattering off D3-branes. It was argued that, due to modular invariance of $\mathcal{N}=4$ SYM, in the large-$N$ limit with Yang-Mills coupling fixed (so-called ``very strong coupling limit") $\zeta(2)/\lambda$ should be promoted to an SL$(2, \mathbb{Z})$ modular function, the non-holomorphic Eisenstein series $E(1; \tau)$.\footnote{Here $\tau= \theta/(2\pi) +  4\pi\,i/g^2_{_{\rm YM}}$. If the result is indeed completed by $E(1; \tau)$, the precise SL$(2, \mathbb{Z})$ completion replaces $\zeta(2)/\lambda$ by $E(1; \tau)/(8N)$; this is determined by using $\lambda=N g^2_{_{\rm YM}}$ and then matching one of the zero modes of $E(1; \tau)$ with $\zeta(2)/\lambda$. Moreover, the first correction given by $E(1; \tau)$ appears to be rather universal, which also arises in other setups of gravitons scattering off D-branes~\cite{Chester:2025ssu, DeLillo:2025stg}.}  Since the EEC should likewise be modular invariant, this similarity suggests the natural conjecture that $\zeta(2)/\lambda$ in  \eqref{eq:leading corr} is also promoted to $E(1; \tau)$. It would be interesting to understand this better and to verify it numerically through the bootstrap with finite Yang-Mills coupling~\cite{Dempsey:2025yiv}. 

\subsection{First curvature correction from $\langle \mathcal{O}_p\,\mathcal{O}_2\,\mathcal{O}_2\,\mathcal{O}_p\rangle$}
Recall that the large-curvature expansion of the $A(S,T)$ can be written order-by-order as a worldsheet integral as defined in (\ref{eq: AdS VS Worldsheet Integral}). Following the bootstrap method detailed in~\cite{Fardelli:2023fyq} we write the following ansatz for $\langle \mathcal{O}_p\,\mathcal{O}_2\,\mathcal{O}_2\,\mathcal{O}_p\rangle$ at first curvature
\begin{equation}
\label{eq:AdS VS KK Ansatz}
    A^{(1)}(S, T) = B^{(1)}(S, T) + B^{(1)}(S, U) + B^{(1)}(U, T) + C^{(1)}(S, T) + C^{(1)}(S, U)\, ,
\end{equation}
which is explicitly constructed to respect the crossing symmetry $A^{(1)}(S, T) = A^{(1)}(S, U)$, $U = -S-T$. The two functions $B^{(1)}(S, T)$ and $C^{(1)}(S, T)$ are expressed as integrals of worldsheet correlators which we write as $G$ and $\tilde{G}$ respectively, i.e.
\begin{align}
    &B^{(1)}(S, T) = \int \d^2z |z|^{-2S-2} |1-z|^{-2T-2} \, G^{(1)}(z, \bar z; S,T)\, , \cr
    &C^{(1)}(S, T) = \int \d^2z |z|^{-2S-2} |1-z|^{-2T-2} \, \tilde{G}^{(1)}(z, \bar z; S,T)\, ,
\end{align}
and we enforce $C^{(1)}(S, T) = 0$ when $p=2$ since the amplitude should be permutation symmetric in this special case. As shown in~\cite{Fardelli:2023fyq}, these worldsheet correlators admit an explicit solution in terms of single-valued multiple polylogarithms given in Appendix~\ref{app: Worldsheet correlators}. 

When evaluating $I^{(1)}(T)$ we will separate the integrals into two categories according to the discussion in subsection \ref{sec:method}. 
We obtain three integrals over the unit disk and three on the unit circle, which we will refer to as bulk and boundary contributions respectively. The detailed evaluation of these integrals are given in Appendix~\ref{app: First Curvature Worldsheet}, and the final result for $I^{(1)}(T)$ is given as 
\begin{align}
\label{eq: p22p I1 full}
    I^{(1)}(T) =& \frac{2\zeta(3) \Gamma({-}2T{-}1)}{\Gamma({-}T)^2} - \frac{p^2 T}{18} \frac{\partial^3}{\partial T^3} \frac{\Gamma({-}2T)}{\Gamma(1{-}T)^2} + \frac{p(p-2)T}{12} \frac{\partial^2}{\partial T^2} \frac{\Gamma(1{-}2T)}{T^2\Gamma(1{-}T)^2} \nonumber \\ 
    & - \frac{(p^2+6)}{36} \frac{\partial^3}{\partial T^3} \frac{\Gamma(1{-}2T)}{\Gamma(1{-}T)^2} + F(T) - F(T-1) \, , 
\end{align}
where
\begin{equation}
    F(T) = \sum_{n=1}^\infty \frac{2\sin(\pi T) \Gamma({-}2T{-}1) \Gamma(n{+}T{+}1)}{\pi n^3 \Gamma(n{-}T)} \, .
\end{equation}

As before, the EEC coefficients in strong coupling expansion are found by first expanding $I^{(1)}(T)$ at small $T$. We find 
\begin{multline}
\label{eq:p22p first curvature}
    I^{(1)}(T) = \zeta(2) - \frac{1}{3}(p^2 + 6) \zeta(3) \cr + T \, \Bigl(-2\zeta(2) + 4\zeta(3) - \frac{1}{12} (19p^2 +19p + 204)\zeta(4) \Bigr) + \mathcal{O}(T^2)\, ,
\end{multline}
where we have dropped the singular term, which corresponds to supergravity contribution that we omit. The $\mathcal{O}(T^0)$ term of \eqref{eq:p22p first curvature} determines $c_{1,1}^{(1)}$, which we can read off using \eqref{eq:expandI} and \eqref{eq:cknseparate}.
Combining with the flat-space term $c_{1,1}^{(0)}$ from \eqref{eq: c11 from flat space worldsheet} we arrive at
\begin{equation}
\label{c11 p22p from worldsheet}
    c_{1,1} = \frac{\Gamma(p+3)}{\Gamma(p-1)\Gamma(p)} \bigl(\zeta(2) - (p^2+2p+2) \zeta(3)\bigr)\, .
\end{equation}
This, alongside the coefficients $c_{0,0}$ and $c_{1,0}$ from \eqref{eq: c00 p22p} and \eqref{eq: Flat space coeffs p22p from worldsheet} respectively, and the functions $Q_{k-n,p}(\xi)$ from \eqref{eq:Qnp}, determines the EEC associated with operators $\mathcal{O}_p$ up to $\mathcal{O}(\lambda^{-3/2})$:
\begin{equation}
\label{eq: EEC p22p from ws}
    \operatorname{EEC}_p(\xi) = 1 + \frac{c_{0,0}}{\lambda} Q_{0,p}(\xi) + \frac{c_{1,1} \, Q_{0,p}(\xi) + c_{1,0} \, Q_{1,p}(\xi)}{\lambda^{3/2}} + \mathcal{O}(\lambda^{-2})\, .
\end{equation}
In the special case of $p=2$, it recovers the result in~\cite{Ren:2026zxs}.

We finish this subsection by computing $c_{2,0}$ and $c_{2,1}$, as they are relevant for the EEC at order $\mathcal{O}(\lambda^{-2})$, which we will consider next. In particular, $c_{2,0}$ is determined by the flat-space contribution and is given in \eqref{eq: Flat space coeffs p22p from worldsheet}. And from \eqref{eq:relationcandd}, we see that $c_{2,1}$ consists of a flat-space contribution from the $\mathcal{O}(t/\lambda)$ term in \eqref{eq: flat space worldsheet integral small T} and a first curvature contribution from the $\mathcal{O}(t^0/\sqrt{\lambda})$ term in \eqref{eq:p22p first curvature}. Putting these two terms together, we obtain 
\begin{equation}
\label{eq: p22p c21}
    c_{2,1} = \frac{\Gamma(p+4)}{\Gamma(p-1) \Gamma(p)} \left(-\zeta(2) + 2\zeta(3) -\frac18 \left(19p^2 +57p +68\right)\zeta(4) \right) \, .
\end{equation}

\subsection{Second curvature correction from $\langle \mathcal{O}_2\,\mathcal{O}_2\,\mathcal{O}_2\,\mathcal{O}_2\rangle$}
\label{sec: Second Curvature From Worldsheet}

We now turn to the second curvature correction, for which the worldsheet correlator has been bootstrapped at $p=2$~\cite{Alday:2023mvu}. At this order the function $I^{(2)}(T)$ is given by the crossing-symmetric worldsheet integral
\begin{equation}
    I^{(2)}(T) = \int_{-i \infty}^{i \infty} \frac{\d S}{2\pi i} \int \d^2z \, |z|^{-2S-2} \, |1-z|^{-2T-2} \, G^{(2)}_{\text{tot}}(z, \bar z; S, T)\, ,
\end{equation}
where $G^{(2)}_{\text{tot}}$ is a linear combination of single-valued multiple polylogarithms of weight six \cite{Alday:2023mvu}, with rational coefficients in $S$ and $T$, whose explicit expansion coefficients are given in Appendix~\ref{app: Worldsheet correlators}.

For the EEC at $\mathcal{O}(\lambda^{-2})$, we only require the $\mathcal{O}(T^0)$ term of $I^{(2)}(T)$: according to \eqref{eq:cknseparate}, the higher powers of $T$ in $I^{(2)}(T)$ contribute only at $\mathcal{O}(\lambda^{-5/2})$ and beyond. This considerably simplifies the computation in two ways. Firstly, the integrals over the unit disk (of the form~\eqref{eq:bulkgeneral}) do not contribute to $I^{(2)}(T)$ at order $\mathcal{O}(T^{0})$.  
As argued in Appendix~\ref{app: Second Curvature Worldsheet}, the disk integrals behave as $T^3 \bigl(\partial_T^6 + \partial_T\bigr) T^{-1} \sim T^{-4} + T$, where the singular powers of $T$ correspond to the supergravity contribution and the stringy corrections start at order $\mathcal{O}(T)$. 
Secondly, for the integrals on the unit circle (of the form~\eqref{eq:boundarygeneral}), we only need to keep the terms of $G^{(2)}_{\text{tot}}$ carrying non-positive powers of $T$. This is because the worldsheet integrals on the unit circle themselves always have finite $T \to 0$ limits: a divergence could only arise from the point $z=1$ on the contour, however such endpoint contributions always produce ratios of Gamma functions of the form $\Gamma(-2T)/\Gamma(-T)^{2}|_{T\sim 0}$, which vanish as $T \to 0$. Since the integration cannot generate a pole in $T$, any term with a positive prefactor power of $T$ only contributes at $\mathcal{O}(T)$ or higher, so in the $T \to 0$ limit only the terms carrying non-positive powers of $T$ survive.

The details of computation of these terms relevant at order $\mathcal{O}(T^0)$ are given in Appendix~\ref{app: Second Curvature Worldsheet}, and the final result for $I^{(2)}(T)$ takes the following form,
\begin{equation}
\label{eq: I2 T0 from worldsheet}
    I^{(2)}(T) = 6 \, \zeta(2) - \frac{13}{2}\zeta(3) + \frac{263}{6}\zeta(4) + \zeta(2)\zeta(3) - \frac{5}{3}\zeta(5) + \mathcal{O}(T)\,.
\end{equation} 
According to \eqref{eq:cknseparate}, the above result for $I^{(2)}(T)$ fully determines $c^{(2)}_{2,2}$, and combining with the flat-space and first-curvature contributions from \eqref{eq: flat space worldsheet integral small T} and \eqref{eq:p22p first curvature} respectively gives
\begin{equation}
    c_{2,2} = 304\, \zeta(2) - 476\, \zeta(3) + 24\, \zeta(2)\zeta(3) + 4692\, \zeta(4) - 40\, \zeta(5)\, .
\end{equation}
Together with $c_{2,0}$ from \eqref{eq: Flat space coeffs p22p from worldsheet} and $c_{2,1}$ from \eqref{eq: p22p c21}, this completes the EEC up to $\mathcal{O}(\lambda^{-2})$: 
\begin{align} \label{eq:EEC2-second-cur}
    \operatorname{EEC}_2(\xi) = 1 & + \frac{c_{0,0}}{\lambda} Q_{0,2}(\xi) + \frac{c_{1,1} \, Q_{0,2}(\xi) + c_{1,0} \, Q_{1,2}(\xi)}{\lambda^{3/2}} \nonumber \\
    & + \frac{c_{2,2} \, Q_{0,2}(\xi) + c_{2,1} \, Q_{1,2}(\xi) + c_{2,0} \, Q_{2,2}(\xi)}{\lambda^{2}} + \mathcal{O}(\lambda^{-5/2})\, ,
\end{align}
where we collect all the coefficients below for convenience
\begin{align}
c_{0,0}  &= 24\zeta(2) \, , \quad   c_{1,0} = 120 \zeta(3) \, , \quad  c_{2,0}  = 855 \zeta(4) \, , \nonumber \\
c_{1,1}  &= 24\zeta(2) - 240\zeta(3) \, , \quad c_{2,1}  = - 120\zeta(2) + 240 \zeta(3) - 3870\zeta(4) \, , \\
c_{2,2} &= 304 \zeta(2) - 476 \zeta(3) + 24 \zeta(2)\zeta(3) + 4692 \zeta(4) - 40 \zeta(5)\, .  \nonumber
\end{align}


\section{Energy-energy correlators from the low-energy expansion} \label{sec:low-energy}

An alternative to expressing the EEC as a worldsheet integral is to extract the strong-coupling expansion coefficient of the EEC directly from the low-energy Wilson coefficients of the Mellin amplitude. We present two methods for this: one based on Borel resummation, and the other on analytic continuation and  localisation of Wilson coefficients.

\subsection{Flat space and first curvature correction from $\langle \mathcal{O}_2\,\mathcal{O}_2\,\mathcal{O}_2\,\mathcal{O}_2\rangle$}
Following the conventions in~\cite{Alday:2023mvu}, we write the Mellin amplitude of $\langle \mathcal{O}_2\,\mathcal{O}_2\,\mathcal{O}_2\,\mathcal{O}_2\rangle$ as an expansion around flat space 
\begin{equation}
\label{eq: Mellin amp p=2}
    \hspace{-0.4em} \mathcal{M}_2(s_1, s_2) = \mathcal{M}_\text{\fontsize{7pt}{8pt}\selectfont SUGRA} + \sum_{a, b=0}^\infty \frac{\Gamma(2a {+} 3b {+} 6)}{8^{a{+}b} \lambda^{\frac32 {+} a {+} \frac32 b}} (s_1^2 {+} s_2^2 {+} s_3^2)^a (s_1 s_2 s_3)^b \left(\alpha^{(0)}_{a,b} + \frac{\alpha^{(1)}_{a,b}}{\sqrt{\lambda}} + \cdots\right)\,,
\end{equation}
with $s_1+s_2+s_3=0$ and $s_1 = s-4/3$, $s_2=t-4/3$ when $p=2$. The Wilson coefficients $\alpha^{(k)}_{a,b}$ for $k \geq 1$ represent the AdS curvature corrections to the flat space Virasoro-Shapiro amplitude. 
A procedure for bootstrapping the Wilson coefficients was laid out in~\cite{Alday:2022xwz}, where results are given for $\alpha^{(1)}_{a,b}$ when $b=0,1,2$, for generic $a$. We use these results to determine the EEC for $p=2$ up to $\mathcal{O}(\lambda^{-3/2})$, for which the analytic result in $a$ for   $\alpha^{(1)}_{a,b}$  is critical. 

\subsubsection{Borel resummation}\label{sec: Borel resummation}

Consider the first stringy correction to the EEC, which is determined by the flat space Wilson coefficient $\alpha^{(0)}_{a,0}$. In particular the coefficient of $Q_{0,p=2}(\xi)$ is obtained by Taylor expanding the $b=0$ summand of (\ref{eq: Mellin amp p=2}) to order $\mathcal{O}(s_2^0)$ and performing the following integration,
\begin{equation} \label{eq:c00}
    c_{0,0} = \int_{-i \infty}^{+i \infty} \frac{\d s_1}{2\pi i} \sum_{a=0}^\infty \frac{\Gamma(2a+6)}{4^a}  \alpha^{(0)}_{a,0} \, s_1^{2a} \,,
\end{equation}
where $\alpha^{(0)}_{a,0} = \zeta(2a+3)$, which can be read off from the flat-space Virasoro-Shapiro amplitude. The resulting integral in $s_1$ can be computed using the modified Borel transform
\begin{equation}
\label{eq: Modified Borel Transform}
    \zeta(n) = \frac{2^{n-1}}{\Gamma(n+1)} \int_0^\infty \d\omega \frac{\omega^n}{\sinh^2 (\omega)}.
\end{equation}
We then find
\begin{equation}
\label{eq: c00 from low-energy}
 c_{0,0} =  \int_0^{\infty} \frac{\d \omega}{\sinh^2(\omega)} \int_{-i \infty}^{i \infty} \frac{\d s_1}{2\pi i} \frac{8 \omega^3 \left(3
   s_1^4 \omega^4-9  
   s_1^2 \omega^2 + 10  \right)}{
   \left(s_1^2
   w^2 - 1 \right)^3}  = 24 \, \zeta(2) \, ,
\end{equation} 
which is in agreement with the result obtained from the worldsheet formulation. We also note that the amplitude 
decays as $1/s^2_1$ when $s_1 \to \infty$. We can similarly obtain $c_{1,0} = 120 \zeta(3)$ by considering $\alpha^{(0)}_{a,1}$ at $\mathcal{O}(t)$.

The coefficient $c_{1,1}$ contains contributions from the flat space and first curvature Wilson coefficients, respectively. In particular, we must consider the $\mathcal{O}(t^0)$ term from $\alpha^{(0)}_{a,1}$ and $\alpha^{(1)}_{a,0}$, which leads to: 
\begin{equation}
\label{eq:c11 from low energy}
    c_{1,1} = \int_{-i \infty}^{i \infty} \frac{\d s_1}{2\pi i} \sum_{a=0}^\infty \left(\frac43 \frac{\Gamma(2a+9)}{2^{2a{+}3}} s_1^{2a{+}2} \, \alpha^{(0)}_{a,1} + \frac{\Gamma(2a+6)}{2^{2a}} s_1^{2a} \, \alpha^{(1)}_{a,0}\right).
\end{equation}
Here both Wilson coefficients are given in terms of depth-two multiple zeta values (MZVs), which we define using the decreasing index convention
\begin{equation}
\label{eq:zetasumdef}
    \zeta(a_1, \ldots, a_k) = \sum_{n_1 > \ldots > n_k} \frac{1}{n_1^{a_1} \cdots n_k^{a_k}}\, .
\end{equation}
In particular, 
\begin{align} \alpha^{(0)}_{a,1} = \left(a+\frac{3}{2}\right) \zeta (2a+6)-2 \zeta(2a+5,1) \, , 
\end{align} 
and $\alpha^{(1)}_{a,0}$ is given in equation (3.20) of~\cite{Alday:2022xwz}, which we quote below: 
\begin{align}
\label{eq: alpha1b0}
    \alpha^{(1)}_{a,0}
    &=
    2\zeta(2a+1)\zeta(3)
    -2\zeta(2a+1,3)
    -(2a+1)\zeta(2a+2,2)
    \nonumber\\
    &\quad
    +\frac13\bigl(6a^2+23a+3\bigr)\zeta(2a+3,1)
    -\frac13\bigl(4a^3+16a^2+13a+6\bigr)\zeta(2a+4)\, .
\end{align}
By writing the depth-two MZVs as integral transforms, as we did for single zetas in \eqref{eq: Modified Borel Transform}, we can evaluate $c_{1,1}$. Consider the second term in \eqref{eq:c11 from low energy},
\begin{equation}
\label{eq: first curv int example}
    \mathcal J = \int_{-i\infty}^{+i\infty}\frac{\d s_1}{2\pi i} \sum_{a=0}^{\infty} \frac{\Gamma(2a{+}6)}{4^a} \alpha^{(1)}_{a,0} \, s_1^{2a}\, .
\end{equation}
For simplicity we use the identity $\Gamma(2a+6) = \int_0^\infty \d\rho \, e^{-\rho} \rho^{2a+5}$ such that
\begin{equation}
    \mathcal J = 48 \int_{-i\infty}^{+i\infty}\frac{\d \tilde s}{2\pi i} \sum_{a=0}^{\infty} \alpha^{(1)}_{a,0}\, \tilde s^{2a} \, ,
\end{equation}
where $\tilde s=s_1 \rho/2$. For the depth-one and depth-two multiple zeta values we use the integral representation of MZVs from \eqref{eq:int rep mzvs}\footnote{In this particular case, it turns out $\alpha^{(1)}_{a,0}$ can be expressed in terms of products of zeta-values~\cite{Alday:2022xwz}, and one may apply \eqref{eq: Modified Borel Transform} to perform the Borel resummation.}
\begin{subequations}
    \begin{equation}
    \label{eq: int rep zeta 1}
    \zeta(2a{+}q) = \frac1{\Gamma(2a{+}q)}\int_0^\infty \frac{y_1^{2a+q-1}}{e^{y_1}-1}\d y_1 \, ,
    \end{equation}
    \begin{equation}
    \label{eq: int rep zeta 2}
    \zeta(2a{+}q,w) = \frac1{\Gamma(2a{+}q)\Gamma(w)}\int_{0<y_1<y_2} \frac{y_1^{2a+q-1}}{e^{y_1}-1} \frac{(y_2-y_1)^{w-1}}{e^{y_2}-1}\d y_1\d y_2\, .
    \end{equation}
\end{subequations}
Consider the first two terms from \eqref{eq: alpha1b0}. While $2\zeta(2a+1)\zeta(3)$ and $2\zeta(2a+1,3)$ are individually divergent due to the $\zeta(1)$ singularity at $a=0$, their combination is regular. Utilising the integral representations above, we obtain 
\begin{align}
\label{eq: Borel in example 1}
    &48 \int_{-i\infty}^{+i\infty}\frac{\d \tilde s}{2\pi i} \sum_{a=0}^{\infty} \tilde s^{2a} \bigl(2\zeta(2a+1)\zeta(3) - 2\zeta(2a+1,3)\bigr) \cr 
    =& \, 48 \int_{-i\infty}^{+i\infty}\frac{\d \tilde s}{2\pi i} \int_0^\infty \d y_1 \int_0^\infty \d y_2 \frac{y_2^2 \, e^{y_2} \cosh(\tilde s y_1)}{(e^{y_2}-1)(e^{y_1+y_2}-1)} = 8\pi^2.
\end{align}
Similarly for the third term in \eqref{eq: alpha1b0},
\begin{align}
\label{eq: Borel in example 2}
    &48 \int_{-i\infty}^{+i\infty}\frac{\d \tilde s}{2\pi i} \sum_{a=0}^{\infty} \tilde s^{2a} \bigl(-(2a+1)\zeta(2a+2,2)\bigr) \cr 
    &= -48 \int_{-i\infty}^{+i\infty}\frac{\d \tilde s}{2\pi i} \int_0^\infty \d y_2 \int_0^{y_2} \d y_1 \frac{y_1(y_2-y_1) \cosh(\tilde s y_1)}{(e^{y_1}-1)(e^{y_2}-1)} = -4\pi^2.
\end{align}
A similar calculation applies to the last two terms, and we find 
\begin{align}
\label{eq: borel int example}
    \mathcal J &= 48 \int_{-i\infty}^{+i\infty}\frac{\d \tilde s}{2\pi i}\sum_{a=0}^\infty \alpha^{(1)}_{a,0} \, \tilde s^{2a} 
    = 4\pi^2 - 80\zeta(3)\, .
\end{align}
The flat-space contribution to $c_{1,1}$ is obtained by applying the same method,
\begin{equation} \label{eq:flat-1}
     \int_{-i \infty}^{i \infty} \frac{\d s_1}{2\pi i} \sum_{a=0}^\infty \frac43 \frac{\Gamma(2a+9)}{2^{2a{+}3}} s_1^{2a{+}2} \alpha^{(0)}_{a,1} = -160 \zeta(3)\, .
\end{equation}
Combining this result with \eqref{eq: borel int example}, we obtain $c_{1,1} = 4\pi^2 - 240\zeta(3)$, which matches the worldsheet calculation as given in (\ref{c11 p22p from worldsheet}).

\subsubsection{Analytic continuation and localisation of Wilson coefficients} \label{sec: Analytic specialization} 

To use the Borel resummation method of the previous subsection, one has to apply the necessary integral transform to every zeta value that appears in the Wilson coefficient. Moreover, one has to be careful about grouping terms in order to avoid apparent divergences, for example \eqref{eq: Borel in example 1}. At second curvature the relevant Wilson coefficient contains many more terms and includes higher-depth MZVs (see \eqref{eq: alpha2b0}), and as such the integrations rapidly become difficult. In this subsection we show that the same results can be obtained much more straightforwardly by analytically continuing the parameter $a$ in the coefficients $\alpha^{(k)}_{a, b}$ away from non-negative integers. In particular, the same results can be obtained by localising $a$ in the Wilson coefficients $\alpha^{(k)}_{a, b}$ to a specific \textit{negative} value.

In general, the integrals we encounter have the following form: 
\begin{equation}
    \mathcal{I} = \int_{-i\infty}^{+i\infty} \frac{\d s_1}{2\pi i} \sum_{a=0}^{\infty} f(a) \frac{\Gamma(2a+p)}{4^a} \zeta(2a+q,\vec{w})\, s_{1}^{2a+m} \,,
\end{equation}
with $f(a)$ a finite polynomial of $a$, and $\vec{w}=(w_{1},\dots,w_{n})$ is a list of positive integers and $p$, $q$, $m$ are also integers. In Appendix~\ref{app:ancont}, we show that, using the fact that Mellin amplitudes decay fast enough as $s_1 \to \pm i \infty$~\cite{Caron-Huot:2021rmr}, the $s_1$-integral localises the sum over $a$ at $a = -\frac{m+1}{2}$. The result is the coefficient of $s_1^{2a+m}$, analytically continued in $a$ and evaluated at this point, multiplied by an overall factor of $1/2$:
\begin{equation}\label{eq:localizationresultmain}
        \mathcal{I} = 2^{m}\, f\!\left( -\frac{m+1}{2} \right) \Gamma(p-m-1)\, \zeta(q-m-1,\,\vec{w})\,.
\end{equation}
However, when $q$ is small, the reduced MZV in \eqref{eq:localizationresultmain} can have a non-positive leading argument $q-m-1 \leq 1$, in which case it is divergent and must be regulated. Also detailed in Appendix~\ref{app:ancont}, we introduce a regulator $\varepsilon$ by shifting $m \to m - 2\varepsilon$, expand each term as a Laurent series in $\varepsilon$, and reduce the result to ordinary MZVs via the recursion \eqref{eq:negativeentryrecursion} together with stuffle regularisation \eqref{eq:stufflereg}. The individual terms may carry poles in $\varepsilon$, but after summing over all contributions to a given $\alpha_{a,b}^{(k)}$ the $\varepsilon$-divergences cancel, leaving a well-defined and finite $\mathcal{O}(\varepsilon^0)$ result that is a linear combination of ordinary MZVs.

We will demonstrate the method by evaluating $c_{0,0}$ and $c_{1,1}$ as we did using Borel resummation in the previous subsection. Applying \eqref{eq:localizationresultmain} to $c_{0,0}$ defined in \eqref{eq:c00}, we obtain
\begin{align}
    c_{0,0} = \Gamma(6-1)\, \zeta(3-1) =24\, \zeta(2) \, ,
\end{align}
which is obtained by setting $a=-1/2$ in \eqref{eq:c00} and multiplying by a factor of $1/2$. This produces the same result as \eqref{eq: c00 from low-energy}.

We can apply the same procedure to obtain $c_{1,1}$. Begin with the flat-space contribution, from the left-hand side of \eqref{eq:flat-1}, we immediately obtain the result
\begin{align}
 \frac2 3 \,  \Gamma(6)\, \alpha^{(0)}_{a,1} \vert_{a\to -\frac 32} = -160\, \zeta(3) \, , 
\end{align}
where we have used $\zeta(2,1) = \zeta(3)$. For the first-curvature contribution, \eqref{eq: first curv int example}, we have 
\begin{align}
    \mathcal J = \Gamma(5) \, \alpha^{(1)}_{a,0} \, \Big|_{a \to  - \frac{1}{2}} \, .
\end{align}
Using the expression given in \eqref{eq: alpha1b0}, it is straightforward to compute $\alpha^{(1)}_{a,0} \, \vert_{a \to - \frac{1}{2}}$. The only term requiring care is $-(2a+1)\zeta(2a+2,2)$, for which it is necessary to introduce a regulator $\varepsilon$,
\begin{align} \label{eq:epsilonterm}
    \lim_{\varepsilon \to 0} \, -(2a+1)\zeta(2a+2,2) \vert_{a \to - \frac{1}{2} + \frac{\varepsilon}{2}} =\lim_{\varepsilon \to 0} \, -\varepsilon \, \zeta(1+\varepsilon, 2) = -\zeta(2) \, .
\end{align}
With this result, we have
\begin{align}
    \alpha^{(1)}_{a,0} \, \Big|_{a \to  - \frac{1}{2}} =  2\zeta(0)\zeta(3)
    -2\zeta(0,3)
    -\zeta(2)
-\frac73\zeta(2,1)
    -\zeta(3) \, .
\end{align}
To further simplify the expression, we may use Euler reflection formula to express $\zeta(0,3) = -\zeta(3,0) + \zeta(0)\zeta(3)-\zeta(3)$, and using $\zeta(3,0) = \zeta(2)-\zeta(3)$, we arrive at 
\begin{align} \label{eq:first-curvature-one}
  \Gamma(5)\,   \alpha^{(1)}_{a,0}\, \Big|_{a = - \frac{1}{2}} = 4\pi^2 -80\zeta(3)\, . 
\end{align}
 The result is in agreement with \eqref{eq: borel int example}. 

We note that this result differs from the first curvature correction obtained in \cite{Dempsey:2025yiv}, where the result is given by $8\pi^2 -80\zeta(3)$. This discrepancy, $4\pi^2$, originates from the treatment of the subtle contribution in \eqref{eq:epsilonterm}, which is included in our result but not in theirs. Physically, this reflects the fact that the light-ray transform associated with the $s$-integral and the spectral summation may not commute. In this sense, we always perform the spectral summation first; this is precisely why the Wilson coefficients are expressed in terms of MZVs as defined in \eqref{eq:MZVs}, where the summation over $\delta$ that is related to the spectral summation has already been performed. 

The examples in this subsection illustrate how the method of analytic continuation and localisation of Wilson coefficients reduces the calculation of the EEC 
to a single step: Borel resummation and integration are both absorbed into \eqref{eq:localizationresultmain}, supplemented by the regularisation \eqref{eq:negativeentryrecursion}. Indeed, we will again encounter subtle contributions like the one in \eqref{eq:epsilonterm}; in some cases individual terms in the Wilson coefficients are divergent in the limit, but these divergences do cancel, and one obtains a well-defined result in agreement with the worldsheet formalism. Physically, this method allows one to directly connect Wilson coefficients with the EEC. This simplification becomes increasingly significant in the more involved cases considered in the following sections.

\subsection{Flat-space and first curvature correction from $\langle \mathcal{O}_p\,\mathcal{O}_2\,\mathcal{O}_2\,\mathcal{O}_p\rangle$}
Unlike the $p=2$ case, where crossing symmetry meant that no odd powers of $s_1$ appeared in the Mellin amplitude \eqref{eq: Mellin amp p=2}, for $p > 2$ we must generalise the stringy part of the amplitude to include all positive integer powers of $s_1$ 
\begin{multline}
\label{eq: alt Mellin amp}
    M_p(s_1,s_2) = M_\text{\fontsize{7pt}{8pt}\selectfont SUGRA} + \sum_{b=0}^\infty \sum_{a=b}^\infty \frac{\Gamma(a+b+p+4)}{\Gamma(b+1) \Gamma(p)\Gamma(p-1)} s_1^{a-b} (s_2 s_3)^b \lambda^{-\frac{3}{2}-\frac{a}{2}-\frac{b}{2}} \cr \times \left(\omega^{(0)}_{a,b} + \frac{\omega^{(1)}_{a,b}(p)}{\sqrt{\lambda}} + \cdots\right)\, ,
\end{multline}
where now $s_1 = s- 2p/3$, $s_2 = t - 2p/3$ with $s_3 = -s_1 - s_2$, and we note that the flat-space contribution $\omega^{(0)}_{a,b}$ does not depend on $p$. The low-energy expansion can be conveniently separated into even and odd powers of $s_1$, and correspondingly the Wilson coefficient should be understood as an even and an odd piece
\begin{equation}
    \omega^{(k)}_{a,0}(p) = \begin{cases} 
      \omega^{(k, \text{even})}_{a,0}(p) & a \: ~ \text{even}\, , \\
      \omega^{(k, \text{odd})}_{a,0}(p) & a \: ~ \text{odd}\, ,
   \end{cases}
\end{equation}
where $\omega^{(k, \text{even/odd})}_{a,0}(p)$ are analytic in $a$.
For example, solving the flat-space dispersive sum rules from equation (4.1) of~\cite{Fardelli:2023fyq} gives \begin{equation}
    \omega^{(0)}_{a,0} = (-1)^{-a} \, 2^{-a-1} \left((-1)^a+1\right) \zeta (a+3)\, .
\end{equation}
We see that $\omega^{(0, \text{odd})}_{a,0} = 0$, and 
\begin{equation}
\label{eq: ome to alp}
    2^{2a} \omega^{(0, \text{even})}_{2a,0} = \alpha^{(0)}_{a,0}\, ,
\end{equation}
where $\alpha^{(0)}_{a,0}$ is the Wilson coefficient of the $p=2$ Mellin amplitude \eqref{eq: Mellin amp p=2}. To determine $c_{0,0}$ we then proceed in the same way as for the $p=2$ case in \eqref{eq:c00}:
\begin{align}
\label{eq:c00 gen p}
    c_{0,0} (p)= \frac{1}{\Gamma(p-1)\Gamma(p)} \int_{-i \infty}^{+i \infty} \frac{\d s_1}{2\pi i} \sum_{a=0}^\infty \Gamma(a+p+4) \, \omega^{(0)}_{a,0} \, s_1^a\, . 
\end{align}
Since $\omega^{(0, \text{odd})}_{a,0} = 0$ we may replace $a\to2a$. We can then apply \eqref{eq: ome to alp} to write the integrand in the same form as the $p=2$ calculation from \eqref{eq:c00}. It follows immediately that
\begin{equation}
\label{eq: c00 p22p from low energy}
    c_{0,0} (p)= \frac{\Gamma(p+3)}{\Gamma(p-1) \Gamma(p)} \zeta(2)\, .
\end{equation}
The remaining flat-space contributions $c_{k,0}$ can also be related to the $p=2$ coefficients using the same procedure:
\begin{equation}
\label{eq: flat space coeff from low energy}
    c_{1,0}(p) = \frac{\Gamma(p+4)}{\Gamma(p{-}1) \Gamma(p)} \zeta(3)\, , \qquad c_{2,0} (p)= \frac{19\, \Gamma(p+5)}{16 \, \Gamma(p{-}1) \Gamma(p)} \zeta(4)\, .
\end{equation}

To determine the EEC up to $\mathcal{O}(\lambda^{-3/2})$ we need to determine $c_{1,1} (p) = c^{(0)}_{1,1} (p)+ c^{(1)}_{1,1}(p)$, where once again we have used \eqref{eq:cknseparate} to split $c_{1,1}(p)$ into the contributions coming from flat space and first curvature respectively. For the flat space part, the relevant sum rule is again equation (4.1) of~\cite{Fardelli:2023fyq}, from which we find
\begin{equation}
    \omega^{(0, \text{even})}_{a,1} = 2^{-a-2} ((a+1) \zeta (a+4)-4 \zeta(a+3,1)), \quad \omega^{(0, \text{odd})}_{a,1} = -2^{-a-2} (a+1) \zeta (a+4).
\end{equation}
Expanding to $\mathcal{O}(s_2)$ order and using $s_2 = t-2p/3$, we have
\begin{equation}
    c^{(0)}_{1,1} (p)  = - \frac{2p}{3} \int_{-i \infty}^{+i \infty} \frac{\d s_1}{2\pi i} \sum_{a=0}^\infty \frac{\Gamma(a+5+p)}{\Gamma(p{-}1) \Gamma(p)} \omega^{(0)}_{a,1}\, s_1^a\, .
\end{equation}
Applying the procedure of subsection \ref{sec: Analytic specialization}, $c^{(0)}_{1,1}$ can be obtained by setting $a\to-1$ in the Wilson coefficients: 
\begin{equation}
\label{eq:ome0b1 ac}
    \omega^{(0, \text{even})}_{a,1}\big|_{a \to -1} = -2 \zeta(2, 1) = -2\zeta(3)\,, \quad \omega^{(0, \text{odd})}_{a,1}\big|_{a \to -1} = 0 \, ,
\end{equation}
and it leads to
\begin{align}
\label{eq:flatpart c011 p22p}
    c^{(0)}_{1,1} (p)  = -\frac{2 \, p \, \Gamma (p+4)}{3\Gamma (p{-}1) \Gamma (p)} \zeta(3)\, . 
\end{align}
To obtain $c^{(1)}_{1,1}(p)$, it requires the first curvature Wilson coefficient $\omega^{(1)}_{a,0}(p)$, which is not available in the literature. By following the bootstrap procedure detailed in Appendix~\ref{app: Bootstrap Wilson Coefficients}, we obtain the analytic expressions 
\begin{subequations}
    \begin{align}
    \label{eq: even spin Wilson coeff p22p}
        \omega^{(1, \text{even})}_{a,0}(p) =& \frac{1}{2^{a+2}} \frac13 \Bigl(6 a^2 \zeta(a+3,1)-4 p^2 \zeta(a+3,1)+8a p \, \zeta(a+3,1)\cr &+8p \, \zeta(a+3,1) +30 a \, \zeta(a+3,1)-24 \zeta(a+1,3) \cr &-12 (a+1) \zeta(a+2,2) +12 \zeta(a+3,1)-2 a^3 \zeta (a+4) -2 a^2 p \, \zeta (a+4) \cr &-12 a^2 \zeta (a+4) +a p^2 \zeta (a+4) +p^2 \zeta (a+4) -4 a p \, \zeta (a+4)\cr &-2 p \, \zeta (a+4) -22 a \, \zeta (a+4) +24 \zeta (3) \, \zeta (a+1)-24 \zeta (a+4)\Bigr)\, ,
    \end{align}
    and
    \begin{equation}
    \label{eq: odd spin Wilson coeff p22p}
        \omega^{(1, \text{odd})}_{a,0}(p) = \frac{1}{2^{a+2}} (a+1) \, p \, (p-2) \zeta (a+4)\, .
    \end{equation}
\end{subequations}
As expected \eqref{eq: odd spin Wilson coeff p22p} vanishes when $p=2$ and, much like we found for flat space in \eqref{eq: ome to alp}, the even part $\omega^{(1, \text{even})}_{a,0}(p)$ can be related to the $p=2$ Wilson coefficient via
\begin{equation}
\label{eq: p=2 even Wilson coeff reduced}
    2^{2a} \, \omega^{(1, \text{even})}_{2a,0}(2) = \alpha^{(1)}_{a,0}\, .
\end{equation}
Localising $\omega^{(1)}_{a,0}(p)$ to $a\to-1$ we see 
that $\omega^{(1, \text{odd})}_{a,0}(p) \to 0$, and
\begin{equation}
    \lim_{\varepsilon \to 0} \omega^{(1, \text{even})}_{a,0}(p)\Big|_{a\to-1+\varepsilon} = 2\zeta(2) - \frac23 (p^2 + 6) \zeta(3)\, ,
\end{equation}
where once again we have used \eqref{eq:epsilonterm} for the term $(a+1) \zeta(a+2,2)$. 
Put all the factors together, we obtain
\begin{align}
\label{eq:1cuvpart c011 p22p}
    c^{(1)}_{1,1} (p) = \int_{-i \infty}^{+i \infty} \frac{\d s_1}{2\pi i} \sum_{2a=0}^\infty \frac{\Gamma(a{+}5{+}p)}{\Gamma(p{-}1)\Gamma(p)} \omega^{(1, \text{even})}_{a,0} s_1^a \cr 
    = \frac{\Gamma(3+p)}{\Gamma(p{-}1)\Gamma(p)} \left(\zeta(2) - \frac13 (p^2 + 6) \zeta(3)\right).
\end{align}
Combining with the flat-space result from \eqref{eq:flatpart c011 p22p}, we arrive at
\begin{equation}
    c_{1,1} (p)= \frac{\Gamma(3+p)}{\Gamma(p{-}1)\Gamma(p)} \left(\zeta(2) - (p^2 + 2p + 2) \zeta(3)\right).
\end{equation}
This coefficient, alongside the functions $Q_{0,p}(\xi)$ from \eqref{eq:Qnp} and $c_{0,0}, c_{1,0}$ from \eqref{eq: c00 p22p from low energy}, \eqref{eq: flat space coeff from low energy}, determines the EEC up to $\mathcal{O}(\lambda^{-3/2})$, and precisely agrees with the results from the worldsheet calculation in \eqref{eq: EEC p22p from ws}. It is worth mentioning that the odd-$a$ part of the Wilson coefficient has vanished each time after localising $a$. This behavior is coincidental, and at second curvature order, which we will discuss next, we expect the odd-$a$ part to contribute.

\subsection{Second curvature correction from $\langle \mathcal{O}_2\, \mathcal{O}_2\, \mathcal{O}_2\, \mathcal{O}_2\rangle$}
\label{sec: Second Curvature Low Energy}

The $\mathcal{O}(\lambda^{-2})$ correction of the EEC takes the form
\begin{equation}
    \operatorname{EEC}_2(\xi) \Big|_{\mathcal{O}(\lambda^{-2})} = c_{2,2} \, Q_{0,2}(\xi)+c_{2,1} Q_{1,2}(\xi) + c_{2,0}\, Q_{2,2}(\xi)\, .
\end{equation}
The coefficient $c_{2,0}$ is given by setting $p=2$ in \eqref{eq: flat space coeff from low energy}, and so we will now calculate the remaining two coefficients.

There are three non-zero contributions to $c_{2,1}$ 
\begin{multline}
   c_{2,1} = -\int^{i \infty}_{-i \infty} \frac{\d s_1}{2\pi i} \sum_{a=0}^\infty \left(\frac13\frac{\Gamma (12+2a)}{2^{2+2a} } s_1^{2a+4} \alpha^{(0)}_{a,2} + \frac{\Gamma(9+2a)}{2^{3+2a}} s_1^{2a+2} \alpha^{(1)}_{a,1} \right.\cr\left. + \frac13 \frac{a(a{+}1) \Gamma(6+2a)}{2^{2a-2}} s_1^{2a-2} \alpha^{(0)}_{a,0} \right).
\end{multline}
All the Wilson coefficients appearing above are known from~\cite{Alday:2022xwz}. Analytically continuing $a$ term-by-term to isolate the $s_1$ pole in each case, the Wilson coefficients take the values
\begin{equation}
\label{eq: 2ndcuv ac coeff values}
    \alpha^{(0)}_{a,2}\Big|_{a=-\frac52} = \frac{35}{8} \zeta(4), \quad \lim_{\varepsilon\to0} \alpha^{(1)}_{a,1}\Big|_{a\to -\frac32 + \frac\varepsilon2} = 2\zeta(2) -4\zeta(3) + \frac{53}{2}\zeta(4), \quad \alpha^{(0)}_{a,0}\Big|_{a=\frac12} = \zeta(4),
\end{equation}
and thus
\begin{equation}
    c_{2,1} 
    = -120\zeta(2) + 240\zeta(3) -3870\zeta(4).
\end{equation}

The coefficient $c_{2,2}$ is where the second curvature correction $\alpha^{(2)}_{a,0}$ first appears in the EEC:
\begin{multline}
\label{eq:c22 eqn}
   c_{2,2} = \int^{i \infty}_{-i \infty} \frac{\d s_1}{2\pi i} \sum_{a=0}^\infty \left(\frac19 \frac{\Gamma(12+2a)}{2^{2+2a} } s_1^{2a+4} \alpha^{(0)}_{a,2} + \frac13 \frac{\Gamma(9+2a)}{2^{1+2a} } s_1^{2a+2} \alpha^{(1)}_{a,1} \right.\cr\left. +\frac{\Gamma(6+2a)}{4^a} s_1^{2a} \alpha^{(2)}_{a,0} + \frac19 \frac{a(a+1)\Gamma(6+2a)}{2^{2a-3} } s_1^{2a-2} \alpha^{(0)}_{a,0} \right)\, .
\end{multline}
The analytic expression for $\alpha^{(2)}_{a,0}$ is not available in the literature; in  Appendix~\ref{app: Bootstrap second curv Wilson coeff} we present the bootstrap solution for determining $\alpha^{(2)}_{a,0}$, as given in \eqref{eq: alpha2b0}. The result is given in terms of depth-three MZVs, which in general cannot  be expressed in terms of lower-depth zeta values~\cite{Brown_2020}. The Borel resummation of section~\ref{sec: Borel resummation} would then require a separate integral transform for each higher-depth MZV, and rapidly becomes impractical. However, the analytic method developed in section~\ref{sec: Analytic specialization} applies directly: by \eqref{eq:localizationresultmain} the contribution localises at $a = -\frac12$. 

Directly setting $a = -\frac12$ in $\alpha^{(2)}_{a,0}$ leads to divergent expressions for the individual terms in \eqref{eq: alpha2b0}; more explicitly, this gives MZVs whose leading argument is less than or equal to one. They are evaluated via the regularisation $a = -\frac12 + \frac \varepsilon 2$, as detailed in Appendix~\ref{app:ancont}. The final result is then obtained by a Laurent-expansion around $\varepsilon = 0$, and importantly the $\varepsilon$-poles of the individual terms cancel in the sum, leaving the finite value, which is given as
\begin{equation}
    \lim_{\varepsilon \to 0} \alpha^{(2)}_{a,0} \Big|_{a = -\frac12 + \frac \varepsilon 2 } = 6\, \zeta(2) - \frac{13}{2}\zeta(3) + \frac{263}{6}\zeta(4) + \zeta(2)\zeta(3) - \frac53 \zeta(5)\,.
\end{equation}
Alongside \eqref{eq: 2ndcuv ac coeff values}, we can then evaluate each term in \eqref{eq:c22 eqn} to find
\begin{equation}
    c_{2,2} = 304\, \zeta(2) - 476\, \zeta(3) + 24\, \zeta(2)\zeta(3) + 4692\, \zeta(4) - 40\, \zeta(5)\,,
\end{equation}
which is in precise agreement with the worldsheet computation of section~\ref{sec: Second Curvature From Worldsheet}.

\section{Plots of energy-energy correlators at strong coupling} 
\label{sec:strong-coupling}

We now discuss the angular dependence of the strong-coupling result. We begin with EEC$_2$ for $p=2$, with the result given in \eqref{eq:EEC2-second-cur}.  
In the supergravity limit, the EEC is homogeneous on the detector sphere, reflecting the absence of jet structure at $\lambda=\infty$.  Finite string-length corrections deform this isotropic distribution, and the functions $Q_{n,2}(\xi)$ encode the resulting angular dependence. For $p=2$, our result provides stringy corrections up to the order $\mathcal{O}(\lambda^{-2})$. 
\begin{figure}[t!]
    \centering
    \includegraphics[width=1\linewidth]{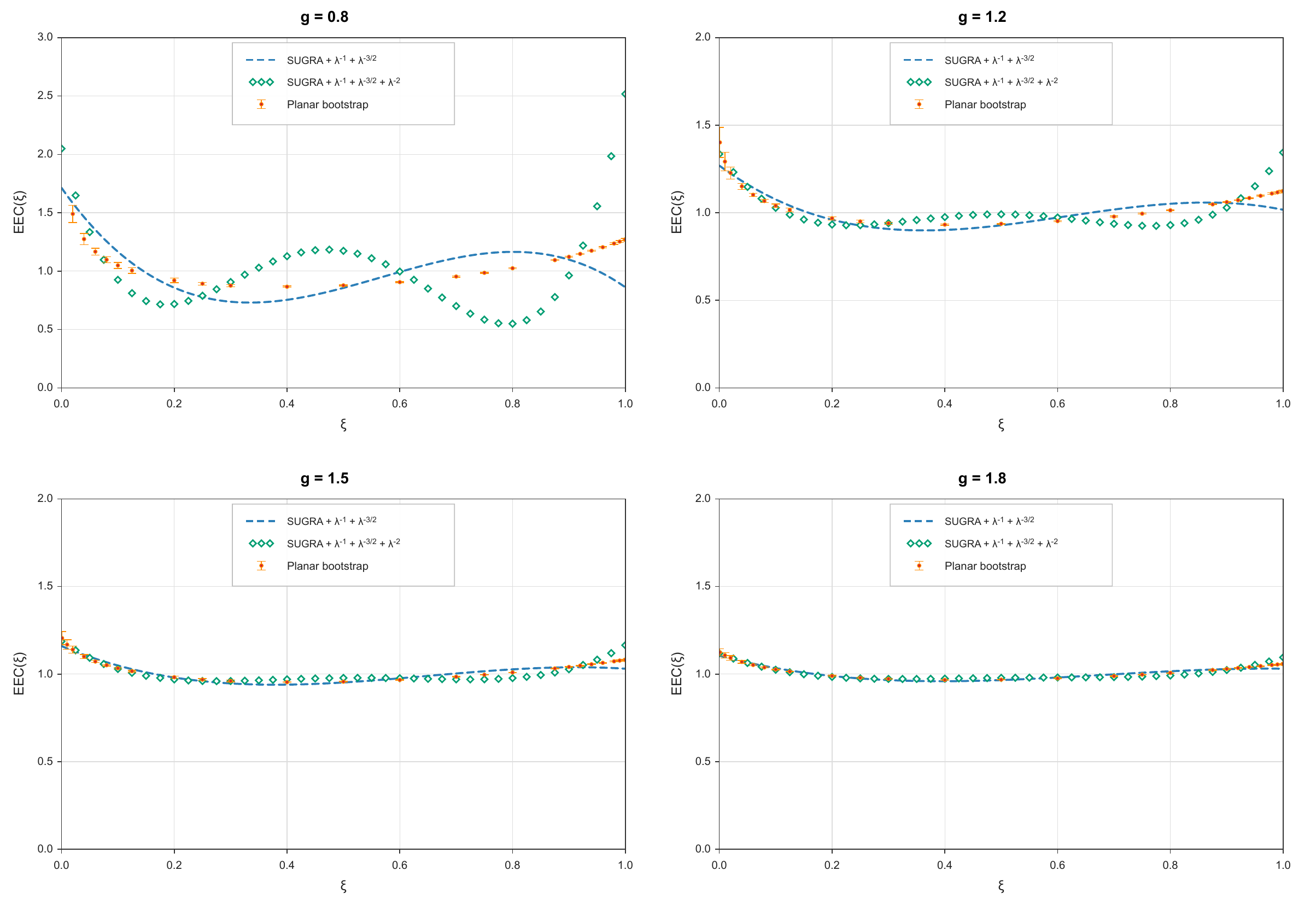}
    \caption{Energy-energy correlator EEC$_2$ at strong coupling. The blue dashed line collects the contributions up to order $\lambda^{-3/2}$ \cite{Ren:2026zxs} while the green hollow diamonds denote the sum through order $\lambda^{-2}$. The red points with yellow error bars show the planar bootstrap data from \cite[fig 14]{Dempsey:2025yiv}.}
    \label{fig:EEC2}
\end{figure}

\Cref{fig:EEC2} compares our strong coupling result with the planar bootstrap data from \cite[fig 14]{Dempsey:2025yiv} at $g=0.8,1.2,1.5,1.8$, where $\lambda=(4\pi g)^2$. The observed trend is consistent with our expectation of the strong-coupling expansion. At the smallest coupling shown, where $g=0.8$, the strong coupling expansion is already rather far from its asymptotic regime, and the deviation from the bootstrap data is correspondingly more pronounced. As $g$ increases, the EEC becomes flatter, approaching the homogeneous supergravity distribution, and the agreement with the planar bootstrap data improves. 

The comparison between the two truncations gives a useful indication of the strong-coupling corrections. The blue dashed curve includes terms through order $\lambda^{-3/2}$, while the green hollow diamonds include the additional $\lambda^{-2}$ contribution. For the larger values of $g$, these two approximations oscillate around the bootstrap data over the central range. This suggests that our curvature expansion is already capturing the dominant behavior in this fixed-angle regime. 

The endpoint regions require a separate treatment. The strong-coupling expansion above should be interpreted at fixed $\xi$ and should not be extrapolated to the strict endpoint limits $\xi\to0$ or $\xi\to1$. In the collinear $\xi\to0$ region, the natural description is the light-ray OPE, where the leading light-ray data determine the endpoint behavior \cite{Hofman:2008ar,Kologlu:2019mfz}. The back-to-back $\xi\to1$ region is sensitive to large logarithms and Sudakov resummation effects \cite{Collins:1981uk,Korchemsky:2019nzm,Chen:2023wah}. In double-logarithmic coordinates, the far end of the back-to-back region exhibits a plateau that is invisible in any fixed-order perturbative expansion. The height of this plateau encodes a nontrivial interplay among operators of different twists and, at finite coupling, develops a nonanalytic dependence on the coupling \cite{Chen:2026lic}. Thus, neither the collinear light-ray OPE behaviour nor the back-to-back Sudakov behavior at the endpoints can be reconstructed from finite strong coupling truncation.
\begin{figure}[t]
   \centering
   \includegraphics[width=1\linewidth]{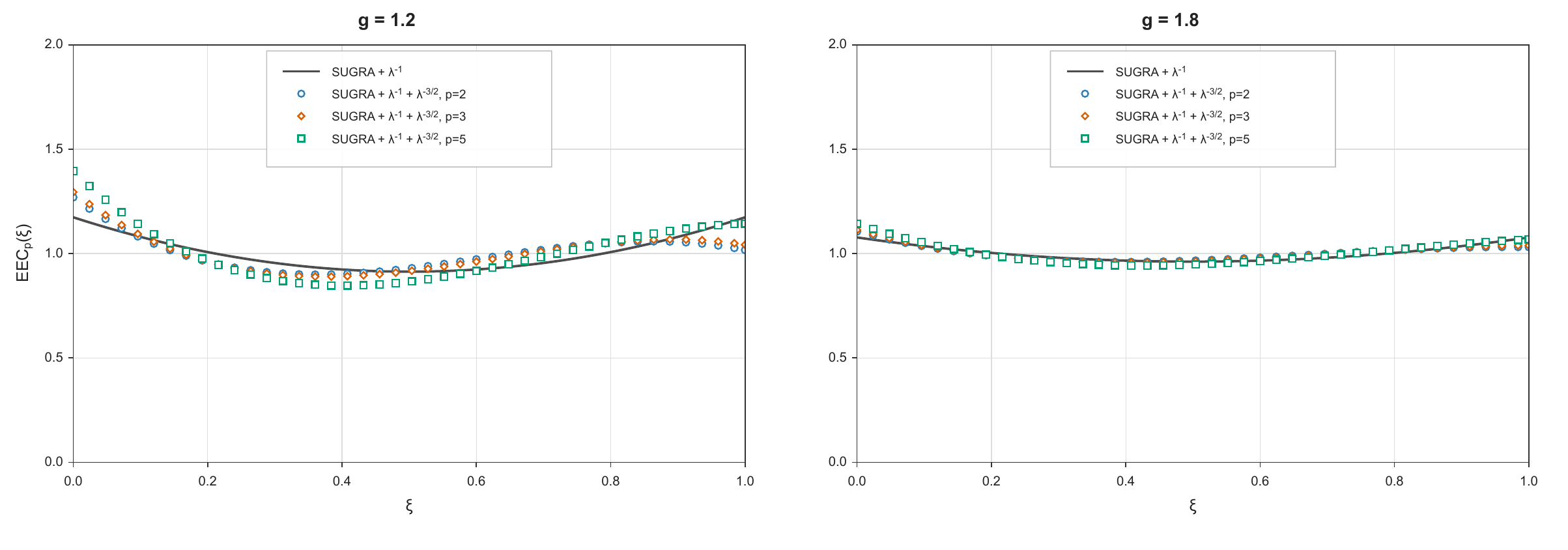}
   \caption{The heavy state energy-energy correlator EEC$_p$ at strong coupling. The black line denote the SUGRA plus $\lambda^{-1}$ correction. The blue hollow circles, red hollow diamonds and green boxes correspond to contributions up to order $\lambda^{-3/2}$ for $p=2,3,5$ separately. }
   \label{fig:EECp}
\end{figure}

Next, we turn to the dependence on the external half-BPS charge $p$, where the first two non-trivial orders for the general EEC$_p$ are derived in \eqref{eq: EEC p22p from ws} with coefficients summarised as: 
\begin{align}
c_{0,0}(p)&=\frac{\Gamma(p{+}3)}{\Gamma(p{-}1)\Gamma(p)}\,\zeta(2)\,,\qquad
c_{1,0}(p)=\frac{\Gamma(p{+}4)}{\Gamma(p{-}1)\Gamma(p)}\,\zeta(3)\,, \nonumber \\ 
c_{1,1}(p)&=\frac{\Gamma(p+3)}{\Gamma(p-1)\Gamma(p)} \bigl(\zeta(2) - (p^2+2p+2) \, \zeta(3)\bigr)\, . 
\end{align}
For general external charge $p$, our result determines the fixed-angle EEC through the first curvature order, namely through $\mathcal{O}(\lambda^{-3/2})$. The first stringy correction has a simple universal form: although both the coefficient and the angular function separately depend on $p$, the product $c_{0,0}(p) \,Q_{0,p}(\xi)$ is independent of $p$. This is the reason we use a single $\mathrm{SUGRA}+\lambda^{-1}$ curve in \Cref{fig:EECp}.

The next correction, at order $\lambda^{-3/2}$, is sensitive to the external charge and grows with $p$ at fixed coupling. The figure illustrates this effect for $p=2,3,5$ at $g=1.2$ and $g=1.8$: increasing $p$ enhances the deviation from the universal leading stringy curve, while increasing $g$ suppresses the correction and drives the EEC back toward the homogeneous supergravity distribution.
\begin{figure}[t]
    \centering
    \includegraphics[width=0.8\linewidth]{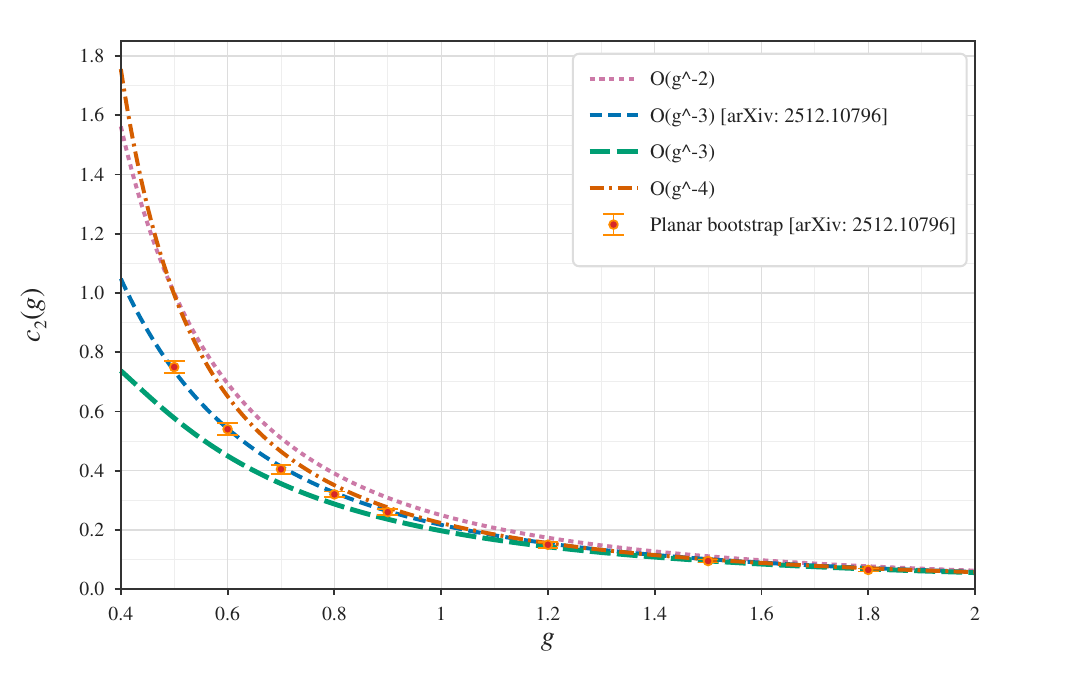}
    \caption{Second EEC multipole $c_2(g)$ at strong coupling compared with numerical bootstrap data. The magenta dotted curve is the leading $\mathcal{O}(g^{-2})$ result; the blue and green dashed curves are the $\mathcal{O}(g^{-3})$ order results of \cite{Dempsey:2025yiv} and our work, respectively; the orange dot-dashed curve includes the new $\mathcal{O}(g^{-4})$ correction from our calculation; and the red points with yellow error bars are the numerical data from planar bootstrap \cite{Dempsey:2025yiv}. }
    \label{fig:EECmultipole}
\end{figure}

Following~\cite{Dempsey:2025yiv}, we can also decompose the EEC into Legendre multipoles. We focus on the $p=2$ case: 
\begin{equation}
    {\rm EEC}_2(\xi)=\sum_{s=0}^{\infty} c_s \, P_s(1-2 \xi)\, , 
\end{equation}
with $c_s=(2s+1)\int_0^1 \d\xi \,P_s(1-2\xi)\,{\rm EEC}_2(\xi)$.  The Ward identities fix $c_0=1$ and $c_1=0$, so that $c_2$ is the first non-trivial multipole. It is straightforward to convert our results in terms of polynomials $Q_{n,p}(\xi)$ into the Legendre basis. In particular, for $c_2$ we find
\begin{align}
c_2(g)
={}&\frac{1}{4g^2}
+\frac{\pi^2-30\zeta(3)}{16\pi^3g^3}
+\frac{\mathcal{C}_4}{256\pi^4g^4}
+\mathcal{O}(g^{-5})\, ,
\label{eq:c2-our-expansion}
\end{align}
where,
\begin{align} \label{eq:4-th order}
    \mathcal{C}_4
={}&\frac{92\pi^2}{3}+\frac{674\pi^4}{105}-236\zeta(3)
+4\pi^2\zeta(3)-40\zeta(5)\,.
\end{align}
The strong coupling expansion of the EEC$_2$ has also been computed in \cite{Dempsey:2025yiv} to the order $\mathcal{O}(g^{-3})$.  Denoting the analytic result of \cite{Dempsey:2025yiv} as $c^{\rm theirs}_2(g)$, the two strong-coupling expansions in the same normalisation are
\begin{align} 
    c^{\rm theirs}_2(g)
={}&\frac{1}{4g^2}
+\frac{\pi^2-15\zeta(3)}{8 \pi^3g^3}
+\mathcal{O}(g^{-4})\, . 
\label{eq:c2-CCB-expansion}
\end{align}
Besides determining the $\mathcal{O}(g^{-4})$ term, we note that $c_2(g)$ and $c^{\rm theirs}_2(g)$  already begin to differ at $\mathcal{O}(g^{-3})$, as commented after \eqref{eq:first-curvature-one}.  

\Cref{fig:EECmultipole} compares our result for $c_2(g)$ with numerical bootstrap data of~\cite{Dempsey:2025yiv}. We see that the truncations of $c_2(g)$ at  $\mathcal{O}(g^{-2})$, $\mathcal{O}(g^{-3})$ and $\mathcal{O}(g^{-4})$ fall alternately on opposite sides of the numerical intervals allowed by the planar bootstrap, with the deviations decreasing as $g$ increases. Moreover, the higher-order truncations are noticeably closer to the numerical bounds when $g$ is not too small, a regime where the strong-coupling expansion is expected to be reliable. \Cref{fig:EECmultipole} also shows the plot for $c^{\rm theirs}_2(g)$ up to order $\mathcal{O}(g^{-3})$ as given in \eqref{eq:c2-CCB-expansion}, which appears close to the numerical bounds even for fairly small values of $g$; however, its fate is unclear if we include higher order corrections --- adding the $\mathcal{O}(g^{-4})$ term $\mathcal{C}_4$ in \eqref{eq:4-th order} from our calculation would in fact push it away from, rather than closer to, the numerical bounds. 

\section{Conclusion and discussion}
\label{sec:con and out}

In this paper we have determined the strong-coupling expansion of the EEC in the planar limit of $\mathcal{N}=4$ SYM for states created by half-BPS operators $\mathcal{O}_p$ of arbitrary dimension $p$. Using the worldsheet representation of the AdS Virasoro-Shapiro amplitude with Kaluza-Klein external states, we obtained the EEC$_p$ up to order $\lambda^{-3/2}$ for generic $p$, and for $p=2$ we went one order further, to $\lambda^{-2}$, by incorporating the second curvature correction to the Virasoro-Shapiro amplitude. In parallel, we developed a complementary method in which the same strong-coupling expansion of the EEC is extracted directly from the low-energy expansion of the Mellin amplitude. The perfect agreement between the worldsheet and Mellin-amplitude approaches, order by order in $1/\sqrt\lambda$, is a non-trivial check on both computations; this also clarifies how the strong-coupling expansions of the EEC and low-energy Wilson coefficients are related by the analytic-continuation procedures of section~\ref{sec:low-energy} and appendix~\ref{app:ancont}. Comparing our results to the numerical bounds from the conformal collider bootstrap~\cite{Dempsey:2025yiv}, we find that the new $\mathcal{O}(\lambda^{-2})$ correction extends the agreement between the analytic strong-coupling expansion and the bootstrap data to smaller values of the coupling. Moreover, we explore the qualitative features of the $p$-dependence of the EEC,  with the notable property that the leading stringy correction is universal and $p$-independent. 

Several directions for future work follow naturally from these results.
A first, immediate task is to complete the second curvature correction for general $p$, along the lines of the $p=2$ computation in section~\ref{sec: Second Curvature From Worldsheet}, which would require bootstrapping the second-curvature worldsheet correlator $G^{(2)}(z,\bar z; S,T)$ with Kaluza-Klein external states \cite{Fa:2026xxx} and evaluating the resulting integrals. This would give access to the $\mathcal{O}(\lambda^{-2})$ term of EEC$_p$ and sharpen the comparison with the bootstrap bounds at finite $p$. The AdS Virasoro-Shapiro amplitude has also been studied in other AdS backgrounds \cite{Chester:2024wnb, Chester:2024esn, Jiang:2025oar,Jiang:2026xnh}; it would be interesting to apply our method and exploit these results to study the EEC in the corresponding dual CFTs. 

A second direction concerns the properties of the EEC under S-duality of $\mathcal{N}=4$ SYM~\cite{Montonen:1977sn}. Since the EEC is built from the same Mellin amplitude that controls the stress-tensor four-point function, one expects its coefficients to organise into non-holomorphic modular functions of the complexified coupling $\tau$. In particular, this modular structure requires considering the ``very strong coupling" limit, i.e. large-$N$ expansion with Yang-Mills coupling fixed, as in the case of correlation functions~\cite{Chester:2019jas, Chester:2020vyz, Dorigoni:2021guq}. As we discussed, it appears possible that the leading stringy correction may be SL$(2, \mathbb{Z})$ completed  by the non-holomorphic Eisenstein series $E(1; \tau)$. It is of great interest to verify the aforementioned non-holomorphic Eisenstein series completion  and more generally understand better the modular properties of the EEC in $\mathcal{N}=4$ SYM. 

Throughout this paper the source operators $\mathcal{O}_p$ were taken to have fixed dimension $p$, so that they correspond to light Kaluza-Klein states in the dual string theory. It would be interesting to instead consider the EEC in states created by heavy (or equivalently large-charge) operators, along the lines of \cite{Chicherin:2023gxt, Cuomo:2025pjp}, as well as heavier states such as giant gravitons with dimension scaling as $N$ \cite{McGreevy:2000cw, Hashimoto:2000zp, Corley:2001zk}, and even heavier states considered in \cite{Paul:2023rka, Brown:2023why, Caetano:2023zwe, Aprile:2024lwy}. The relevant four-point functions are then of heavy-heavy-light-light (HHLL) type. Results for giant graviton HHLL correlators have been obtained in both the weak- and strong-coupling expansions~\cite{Jiang:2019xdz, Jiang:2023uut, Chen:2025yxg, Wu:2025ott, Brown:2026dhy, He:2026ios}, and for certain heavy operators the correlators are known to all orders in the large-charge 't Hooft limit~\cite{Caetano:2023zwe, Brown:2024yvt, Brown:2025cbz}. It would be of interest to exploit these HHLL results to study the EEC for these heavy states.

Finally, it would be interesting to understand the fate of our results in the two kinematic regions that lie outside the reach of the fixed-angle strong-coupling expansion discussed here: the collinear limit $\xi \to 0$, governed by the light-ray OPE~\cite{Hofman:2008ar,Kologlu:2019mfz}, and the back-to-back limit $\xi \to 1$, controlled by Sudakov resummation~\cite{Collins:1981uk,Korchemsky:2019nzm}. Matching the strong-coupling expansion of the EEC onto the data controlling these endpoint regions, order by order in $1/\sqrt\lambda$, would provide a complete analytic picture of the energy-energy correlator in $\mathcal{N}=4$ SYM at strong coupling, complementing the numerical results available from the conformal collider bootstrap.

\section*{Acknowledgements}

%
We thank Hao Chen, Alessandro Georgoudis, Ian Moult,  Silviu Pufu, Alexander Zhiboedov and Hua Xing Zhu for helpful discussions. MJ is supported by a Science and Technology Facilities Council (STFC) studentship. LR is supported by the Royal Society via a Newton International Fellowship. BW is supported by the National Science Foundation of China under Grant No.~124B2095, No.~12175197 and Grant No.~12347103. CW is supported by a Royal Society University Research Fellowship No.~URF$\backslash$R$\backslash$221015 and a STFC Consolidated Grant, ST$\backslash$X00063X$\backslash$1 ``Amplitudes, strings \& duality".

\bibliographystyle{JHEP}

\appendix

\section{The derivation of $Q_{n,p}(\xi)$}
\label{app: p22p Kernel Recurrence Relation}

In this section we derive an exact expression for the functions $Q_{n,p}(\xi)$. First we decompose the detector kernel as
\begin{equation}
    \mathcal{K}_p(t;\xi)=\sum_{k=0}^2 \mathcal{K}_{p,k}(t;\xi)\, ,
\end{equation}
where
\begin{multline}
\label{eq:kernel for general spins2}
    \mathcal{K}_{p,k}(t;\xi) =  {2 \choose k} \frac{(-1)^k \xi^{\frac{t}{2}-k} \pi^2 t (t-2) \Gamma(k{-}\frac{t}{2}) \Gamma(p{-}1) \Gamma(p) \Gamma(p{-}\frac{t}{2})}{32 \sin^2(\frac{\pi t}{2}) \Gamma(p{+}k{-}2{-}\frac{t}{2}) \Gamma(k{-}2{-}\frac{t}{2}) \Gamma(-2{-}\frac{t}{2}) \Gamma(p{-}k{+}3{+}\frac{t}{2})} \\ \times \, _2F_1\Bigl(3-k+\frac{t}{2}, 3-k+\frac{t}{2}, p-k+3+\frac{t}{2}; \xi\Bigr)\, ,
\end{multline}
and define
\begin{equation}
    Q_{n,p}(\xi)=\sum_{k=0}^2 Q_{n,p,k}(\xi),
    \qquad
    Q_{n,p,k}(\xi)
    =
    \int_{-i\infty}^{i\infty}
    \frac{\d t}{2\pi i}\,
    t^n\,\mathcal{K}_{p,k}(t;\xi)\, .
\end{equation}
The functions $\mathcal{K}_{p,k}$ satisfy the following differential relation
\begin{equation}
\label{eq: recurrence relation seed kernels}
    \frac{t}{2}
    \bigl(p+2+\xi\partial_\xi\bigr)
    \mathcal{K}_{p,k}
    =
    \left[
    (1-\xi)\xi^2\partial_\xi^2
    +\xi(k+p-7\xi+3)\partial_\xi
    +\bigl(k(p+2)-9\xi\bigr)
    \right]
    \mathcal{K}_{p,k}\, ,
\end{equation}
which in turn define the recurrence relations for $Q_{n,p,k}(\xi)$. This equation can be inverted by imposing regularity at $\xi=0$. To do this we define the functional
\begin{multline}
\mathcal{F}_{p,k}[f](\xi)
:=
2\,\xi^{-(p+2)}
\int_0^\xi \d u\,u^{p+1}
\left[
(1{-}u)u^2f''(u)
+u(k{+}p{-}7u{+}3)f'(u)
\right.\cr\left.+\bigl(k(p{+}2)-9u\bigr)f(u)
\right]\, ,
\end{multline}
such that $Q_{n+1,p,k}(\xi) = \mathcal{F}_{p,k}\!\left[Q_{n,p,k}\right](\xi)$.

We find a general expression for $Q_{n,p,k}(\xi)$ by first introducing the seed functions
\begin{subequations}
\label{seedofQ}
\begin{align*}
    Q_{0,p,0}(\xi)
    &=
    \frac{144\xi^2 \, \Gamma(p{-}1)}{(p)_7}
    \left(
    900 \xi^2
    +50 \xi (p{-}4)(p{+}6)
    +(p{-}4)(p{-}3)(p{+}5)_2
    \right),
    \\
    Q_{0,p,1}(\xi)
    &=
    \frac{144\xi \, \Gamma(p{-}1)}{(p)_7}
    \left(
    -1800 \xi^3
    -50\xi^2(p{-}12)(p{+}6)
    +2\xi(p{+}5)_2(7p{-}36)
    -(p{-}3)(p{+}4)_3\right),
    \\
    Q_{0,p,2}(\xi)
    &=
    \frac{24 \, \Gamma(p{-}1)}{(p)_7}
    \left(
    5400\xi^4
    -2400 \xi^3(p{+}6)
    +432\xi^2(p{+}5)_2
    -36\xi(p{+}4)_3
    +(p{+}3)_4
    \right) ,
\end{align*}
\end{subequations}
where $(a)_n = \Gamma(n{+}a)/\Gamma(a)$ 
is the Pochhammer symbol. Summing the three seed functions gives the known $n=0$ result
\begin{equation} \label{eq:Q0pxi}
    Q_{0,p}(\xi)
    =
    Q_{0,p,0}(\xi)
    +Q_{0,p,1}(\xi)
    +Q_{0,p,2}(\xi)
    =
    \frac{24\, \Gamma(p-1)}{(p)_3}Q_{0, 2}(\xi)\, ,
\end{equation}
where $Q_{0, 2} (\xi)= 6\xi^2-6\xi+1$. 
Each seed function is a polynomial in $\xi$ of at most degree four, and therefore we can write them in the form
\begin{equation}\label{eq:afromQ}
    Q_{0,p,k}(\xi)
    =
    \frac{\Gamma(p-1)}{(p)_7}
    \sum_{m=0}^4 a_m^{(p,k)}\xi^m.
\end{equation}
By acting on $\xi^m$ with the differential operator appearing in (\ref{eq: recurrence relation seed kernels}) we obtain
\begin{equation}
    \mathcal{F}_{p,k}[\xi^m]
    =
    2(m+k)\xi^m
    -
    2\frac{(m+3)^2}{m+p+3}\xi^{m+1}.
\end{equation}
Iterating $n$ times it follows that
\begin{equation}
    \mathcal{F}_{p,k}^{\,n}[\xi^m]
    =
    \sum_{j=0}^n
    2^n
    \frac{(m+3)_j^2}{(m+p+3)_j}
    \mathcal{S}_{m+k}^{(n,j)}
    \xi^{m+j},
\end{equation}
where
\begin{equation}
\label{eq: Srnj}
    \mathcal{S}_{r}^{(n,j)}
    =
    \sum_{s=0}^j
    \frac{(-1)^s}{s!(j-s)!}
    (r+s)^n.
\end{equation}
Hence, after summing over $k$,
\begin{equation}
\label{eq: Qnp before a def}
    Q_{n,p}(\xi)
    =
    \frac{\Gamma(p-1)}{(p)_7}
    \sum_{k=0}^2
    \sum_{m=0}^4
    a_m^{(p,k)}
    \sum_{j=0}^n
    2^n
    \frac{(m+3)_j^2}{(m+p+3)_j}
    \mathcal{S}_{m+k}^{(n,j)}
    \xi^{m+j} \, .
\end{equation}
By inspecting the form of \eqref{eq: Srnj} we notice that
\begin{equation}
    \mathcal{S}_{r+q}^{(n,j)}
    = \sum_{\ell=0}^q (-1)^\ell {q \choose \ell} (j+1)_\ell \, S_r^{(n, j+\ell)},
\end{equation}
which we can insert into \eqref{eq: Qnp before a def} to obtain
\begin{multline}
    Q_{n,p}(\xi)
    =
    \frac{\Gamma(p-1)}{(p)_7}
    \sum_{k=0}^2
    \sum_{m=0}^4
    a_m^{(p,k)}
    \sum_{j=0}^n
    2^n
    \frac{(m+3)_j^2}{(m+p+3)_j} \xi^{m+j} \cr \times \sum_{\ell=0}^{m+k-2} (-1)^\ell {m+k-2 \choose \ell} (j+1)_\ell \, S_2^{(n, j+\ell)}.
\end{multline}
Since $S_2^{(n,j+\ell)} = 0$ when $j+\ell>n$ we can shift $j \to j-\ell$, explicitly insert the coefficients $a_m^{(p,k)}$ from \eqref{eq:afromQ} and perform the sums over $m$, $k$ and $\ell$ to arrive at the final expression for $ Q_{n,p}(\xi)$, which can be written in the compact form
\begin{equation}
\label{eq:Qnp}
    Q_{n,p}(\xi)
    =
    \frac{12\,2^n\,\Gamma(p-1)}{(p)_3}
    \frac{\mathrm d^2}{\mathrm d\xi^2}
    \left[
    (1-\xi)^2
    \sum_{j=0}^{n}
    \frac{(3)_j(5)_j}{(p+3)_j}
    \mathcal S_2^{(n,j)}\xi^{j+2}
    \right].
\end{equation}
We list the relevant results as explicit examples,
\begin{align}
    Q_{0,p}(\xi) =& \frac{24  \Gamma (p-1)}{(p)_3} \left(6 \xi ^2-6 \xi +1\right) \, ,\\
    Q_{1,p}(\xi) =&\frac{48 \Gamma (p-1)}{(p)_4} \left(-150 \xi ^3+216 \xi ^2-81 \xi +12 \xi ^2 p-12 \xi  p+2 p+6\right) \, , \\
    Q_{2,p}(\xi) =& \frac{96 \Gamma (p-1)}{(p)_5} \left(5400 \xi ^4-10200 \xi ^3+6048 \xi ^2-1188 \xi +24 \xi ^2 p^2-24 \xi  p^2 \right. \nonumber \\
    & \left. +4 p^2-750 \xi ^3 p+1068 \xi ^2 p-393 \xi  p+28 p+48\right)   \, .
\end{align}

\section{Worldsheet correlator and worldsheet integrals} \label{app:worldsheet}

\subsection{Single-valued multiple polylogarithms}
\label{app: SVMPLs}
Harmonic multiple polylogarithms (MPLs) are multi-valued analytic functions $L_w(z)$ labeled by a word $w$. A word is a string of ``letters" taken from the alphabet \{0,1\}, for example $010$. The transcendental weight of $L_w(z)$ is simply the length of the word $w$. MPLs can be defined recursively in the following sense
\begin{equation}
\label{eq:MPLS diff eqs}
    \frac{\partial}{\partial z} L_{0w}(z) = \frac{L_w(z)}{z}, \quad \frac{\partial}{\partial z} L_{1w}(z) = \frac{L_w(z)}{z-1}, \quad L_\emptyset(z)=1\, .
\end{equation}
These functions are multi-valued in general, however, a theorem due to Brown~\cite{Brown_2004, Brown:2013gia} establishes the existence of a unique family of \emph{single-valued} functions $\mathcal{L}_w(z)$ satisfying the same differential equations given in (\ref{eq:MPLS diff eqs}). Moreover each single-valued MPL (SVMPL) can be written explicitly as a linear combination of functions $L_{w'}(z) L_{w''}(\bar{z})$, where $w$ is given by the concatenation of $w'$ and $w''$.

These single-valued functions play crucial roles in constructing AdS Virasoro-Shapiro amplitude \cite{Alday:2023jdk, Alday:2023mvu}, generalising results for flat-space string amplitudes~\cite{Schlotterer:2012ny, Broedel:2013tta}. 
For the first curvature correction  $A^{(1)}(S,T)$, we need a basis of weight-3 SVMPLs:  
\begin{align}
\label{eq: SVMPL first curv basis}
\mathcal{L}_{000}(z) &= L_0(z)L_{00}(\bar z) + L_0(\bar z)L_{00}(z) + L_{000}(z) + L_{000}(\bar z)\, , \nonumber \\
\mathcal{L}_{001}(z) &= L_0(z)L_{10}(\bar z) + L_1(\bar z)L_{00}(z) + L_{001}(z) + L_{100}(\bar z)\, , \nonumber \\
\mathcal{L}_{010}(z) &= L_0(z)L_{01}(\bar z) + L_0(\bar z)L_{01}(z) + L_{010}(z) + L_{010}(\bar z)\, , \nonumber \\
\mathcal{L}_{011}(z) &= L_0(z)L_{11}(\bar z) + L_1(\bar z)L_{01}(z) + L_{011}(z) + L_{110}(\bar z)\, , \nonumber \\
\mathcal{L}_{100}(z) &= L_1(z)L_{00}(\bar z) + L_0(\bar z)L_{10}(z) + L_{100}(z) + L_{001}(\bar z)\, , \nonumber \\
\mathcal{L}_{101}(z) &= L_1(z)L_{10}(\bar z) + L_1(\bar z)L_{10}(z) + L_{101}(z) + L_{101}(\bar z)\, , \nonumber \\
\mathcal{L}_{110}(z) &= L_1(z)L_{01}(\bar z) + L_0(\bar z)L_{11}(z) + L_{110}(z) + L_{011}(\bar z)\, , \nonumber \\
\mathcal{L}_{111}(z) &= L_1(z)L_{11}(\bar z) + L_1(\bar z)L_{11}(z) + L_{111}(z) + L_{111}(\bar z)\, .
\end{align} 
The second curvature correction $A^{(2)}(S,T)$ then requires a basis of SVMPLs with weight six \cite{Alday:2023mvu}.  The Mathematica package PolyLogTools~\cite{Duhr:2019tlz} provides an efficient implementation of these SVMPLs. 

\subsection{Worldsheet correlators}
\label{app: Worldsheet correlators}
Using the ansatz in (\ref{eq:AdS VS KK Ansatz}), the first curvature correction $A^{(1)}(S,T)$ for the correlator $\langle \mathcal{O}_p \mathcal{O}_2 \mathcal{O}_2 \mathcal{O}_p \rangle$ is completely specified by the following two worldsheet correlators~\cite{Fardelli:2023fyq}
\begin{subequations}
    \begin{align}
    \label{eq:p22p worldsheet correlator 1}
        G^{(1)}(z, \bar z; S,T) & = \frac{1}{6}\Big(
        -p^2 \mathcal{L}_{000}
        +p(p-1) \mathcal{L}_{001}
        +(p^2-p-3) \mathcal{L}_{010}\\
      &\hspace{3.5em} -p \mathcal{L}_{011}
        +p(p-1) \mathcal{L}_{100}
        -(p+3) \mathcal{L}_{101}
        -p \mathcal{L}_{110}
        \Big) \cr
       & +\frac{p^2 T}{6(S+T)}
        \Big(
        \mathcal{L}_{000}
        -\mathcal{L}_{001}
        -\mathcal{L}_{010}
        +\mathcal{L}_{011}
        -\mathcal{L}_{100}
        +\mathcal{L}_{101}
        +\mathcal{L}_{110}
        -\mathcal{L}_{111}
        \Big)\, , \notag
    \end{align}
    and
\begin{align} \label{eq:p22p worldsheet correlator 2}
    \tilde{G}^{(1)}(z, \bar z; S,T) = \frac{p (p-2)}{6} \Bigl(3 \mathcal{L}_{000}(z) - 2 \mathcal{L}_{001}(z) - 2\mathcal{L}_{010}(z) - 2\mathcal{L}_{100}(z)\Bigr) \cr + \frac{p(p-2)T}{6(S+T)} \Bigl(-3\mathcal{L}_{000}(z) + 2\mathcal{L}_{001}(z) + 2\mathcal{L}_{010}(z) - \mathcal{L}_{011}(z) \cr+ 2\mathcal{L}_{100}(z) - \mathcal{L}_{101}(z) - \mathcal{L}_{110}(z)\Bigr)\, .
\end{align}
In the case of $\langle \mathcal{O}_2\mathcal{O}_2\mathcal{O}_2\mathcal{O}_2 \rangle$, the second curvature correction of the worldsheet correlator has been determined \cite{Alday:2023mvu} and it reads
\begin{equation}
    G^{(2)}(z,\bar{z};S,T) = \sum\limits_{u} r^{(2)s}_{u} \mathcal{L}^{(2)s}_u +\sum\limits_{v} r^{(2)a}_{v} \mathcal{L}^{(2)a}_v \,,
\end{equation}
where one can split the SVMPLs into symmetric and anti-symmetric components based on $z \to 1-z$,
\begin{align}
\mathcal{L}^s_{w}(z) ={}& \mathcal{L}_{w}(z) + \mathcal{L}_{w}(1-z)+\mathcal{L}_{w}(\bar{z}) + \mathcal{L}_{w}(1-\bar{z})\,,\\
\mathcal{L}^a_{w}(z) ={}& \mathcal{L}_{w}(z) - \mathcal{L}_{w}(1-z)+\mathcal{L}_{w}(\bar{z}) - \mathcal{L}_{w}(1-\bar{z})\,.
\end{align}
The detailed expression of weight 6 basis $\mathcal{L}^{(2)s/a}_w$ is recorded in \cite{Alday:2023mvu}, while the final coefficients are given by
\begin{align}
r^{(2)s} ={}& \frac{S^2 + T^2}{2^4 3^5} \Big(-216,26739,13111,-7271,-9286,-9139,-26100,9219,-12672,-15917,\nonumber\\
&3541,-9901,-823,29697,-17307,1674,10530,3780,23760,-3483,0,0,0,0,0 \Big)\nonumber\\
& + \frac{S T}{2^4 3^5} \Big( 
216,8163,24433,-132845,-33460,-92347,-25725,67200,-21045,18571,\nonumber\\
&27967,-7363,52694,9372,1848,-7575,21006,26760,26769,55233,0,0,0,0,0
\Big)\,,\nonumber\\
r^{(2)a} ={}& \frac{(S^2 + T^2)(S-T)}{2^4 3^5 (S+T)} \Big(
-216,26739,13111,-7271,-9286,-9139,-26100,-12672,\nonumber\\
&-15917,-9901,2417,17061,-17307,1674,-432,3483,0,0,0,0
\Big) \nonumber\\
&+ \frac{ST(S-T)}{2^4 3^5 (S+T)} \Big(
-216,61641,50655,-56985,-52032,4521,-26307,-2387,\nonumber\\
&-21173,5559,-43268,-16916,-67642,-19393,11432,15345,0,0,-84456,-5292
\Big)\,. \nonumber
\end{align}

\end{subequations}

\subsection{Worldsheet integrals at first curvature}
\label{app: First Curvature Worldsheet}
\paragraph{Bulk terms.} Using the ansatz expression for $A^{(1)}(S,T)$ from (\ref{eq:AdS VS KK Ansatz}) we obtain three integrals over the unit disk, which we will label $I^{(1)}_i(T)$, with $i = 1, 2, 5$ to match the conventions of the $p=2$ case from~\cite{Ren:2026zxs}. The first of these comes from $B^{(1)}(S,T)$, in particular from the $1/(S+T)$ term in the worldsheet correlator \eqref{eq:p22p worldsheet correlator 1}
\begin{align}
    I^{(1)}_1(T) = \int \frac{\d S}{2\pi i} \int \d^2z \, |z|^{-2S-2} |1-z|^{-2T-2} \frac{p^2 T}{6(S+T)} 
        \Big(
        \mathcal{L}_{000}
        -\mathcal{L}_{001}
        -\mathcal{L}_{010}
        +\mathcal{L}_{011}
        \cr -\mathcal{L}_{100}
        +\mathcal{L}_{101}
        +\mathcal{L}_{110}
        -\mathcal{L}_{111}
        \Big) \cr
        =\frac{p^2 T}{36} \int_{|z|<1}\d^2z \, |z|^{2T-2} \, |1-z|^{-2T-2} \log^3 \frac{|z|^2}{|1-z|^2} = -\frac{p^2 T}{36}\frac{\partial^3}{\partial T^3} \frac{\Gamma({-}2T)}{\Gamma(1{-}T)^2}\, ,
\end{align}
where between the first and second equalities we have simplified the integral using \eqref{eq:Integral case 1} after shifting the integration variable $S \to S-T$.
The second bulk term comes from $B^{(1)}(U,T)$, and in fact yields the same result
\begin{equation}
    I^{(1)}_2(T) =\frac{p^2 T}{36} \int_{|z|<1}\d^2z \, |1-z|^{-2T-2} \log^3 |1-z|^2 = -\frac{p^2 T}{36}\frac{\partial^3}{\partial T^3} \frac{\Gamma({-}2T)}{\Gamma(1{-}T)^2} \, .
\end{equation}
The third bulk contribution is from $C^{(1)}(S,T)$, for which we find
\begin{align}
\label{eq:intCBulk}
    I^{(1)}_5(T) &= -\frac{p(p-2)T}{12} \int_{|z|<1}\d^2z \, |z|^{2T-2} \, |1-z|^{-2T-2} \, \log|z|^2 \, \log^2\frac{|z|^2}{|1-z|^2}  \nonumber \\
    &= \frac{p(p-2)T}{12} \frac{\partial^2}{\partial T^2}\Bigl(\frac{\Gamma(1-2T)}{T^2 \Gamma(1-T)^2}\Bigr)\, .
\end{align}

\paragraph{Boundary terms.} The remaining parts of $B^{(1)}(S,T)$ and $B^{(1)}(U,T)$ contribute to an integral around the unit circle. Amazingly, all $p$ dependence vanishes and we obtain the integral
\begin{align}
    I^{(1)}_3(T) &= \oint_{|z|=1} \frac{\d\theta}{2\pi} \, |1-z|^{-2T-2} (2\zeta(3) - \operatorname{Li}_3(z) - \operatorname{Li}_3(\bar{z})) \nonumber \\
    &= \frac{2\zeta(3) \Gamma({-}2T{-}1)}{\Gamma(-T)^2} + F(T)\, ,
\end{align}
where we have defined
\begin{equation}
    F(T) = \sum_{n=1}^\infty \frac{2\sin(\pi T) \Gamma({-}2T{-}1) \Gamma(n{+}T{+}1)}{\pi n^3 \Gamma(n{-}T)}\, .
\end{equation}
In contrast, the contribution from $B^{(1)}(S, U)$ lies entirely on the boundary, and simplifies to
\begin{align}
    I^{(1)}_4(T) &= \oint_{|z|=1} \frac{\d\theta}{2\pi} \, |1-z|^{-2T} \Bigl(\operatorname{Li}_3(z) + \operatorname{Li}_3(\bar{z}) - \frac{1}{18}(p^2-3p-3) \log^3|1-z|^2 \Bigr) \nonumber \\
    &= \frac{(p^2{-}3p{-}3)}{18} \frac{\partial^3}{\partial T^3} \frac{\Gamma(1{-}2T)}{\Gamma(1{-}T)^2} - F(T-1)\, .
\end{align}
Finally we have the contribution from $C^{(1)}(S, U)$
\begin{equation}
\label{eq: 1cuv I6}
    I^{(1)}_6(T) = \frac{p(p-2)}{12} \oint_{|z|=1} \frac{\d\theta}{2\pi} \, |1-z|^{-2T} \log^3|1-z|^2 = -\frac{p(p-2)}{12} \frac{\partial^3}{\partial T^3} \frac{\Gamma(1{-}2T)}{\Gamma(1{-}T)^2}\, .
\end{equation}
Collecting all terms $I^{(1)}(T) = \sum_{i=1}^6 I^{(1)}_i(T)$, we arrive at the final result, given in \eqref{eq: p22p I1 full} in the main text.

\subsection{Worldsheet integrals at second curvature}
\label{app: Second Curvature Worldsheet}

The second-curvature worldsheet correlator of Appendix~\ref{app: Worldsheet correlators}, summed over its crossing images, can be uniquely decomposed according to its $S$- and $T$-dependence from \cite[eq. (2.11)]{Alday:2023mvu} as
    \begin{align}
\label{eq: G2 channel decomposition}
       & G^{(2)}_{\text{tot}}(z,\bar z; S,T) = \frac{T^3}{S{+}T}\, G^{ST}_1(z,\bar z) {+} T^2\, G^{ST}_2(z,\bar z) {+} ST\, G^{ST}_3(z,\bar z) {+} S^2\, G^{ST}_4(z,\bar z) \cr
        & + |z|^2 \left( \frac{T^3}{S}\, G^{UT}_1(z,\bar z) + T^2\, G^{UT}_2(z,\bar z) + ST\, G^{UT}_3(z,\bar z) + S^2\, G^{UT}_4(z,\bar z) \right) \cr
        & + |1-z|^2 \left( \frac{S^3}{T}\, G^{SU}_1(z,\bar z) + S^2\, G^{SU}_2(z,\bar z) + ST\, G^{SU}_3(z,\bar z) + T^2\, G^{SU}_4(z,\bar z) \right) ,
    \end{align}
where the $G^{XY}_i(z,\bar z)$ are SVMPLs of weight six, independent of $S$ and $T$. Following the discussion in subsection~\ref{sec:method}, the negative powers of $S$ produce integrals over the unit disk, while the non-negative powers localise on the unit circle,
\begin{equation}
    I^{(2)}(T) = I^{(2)}_{\text{disk}}(T) + I^{(2)}_{\text{circ}}(T)\, .
\end{equation}

\paragraph{Bulk terms.} The simple poles in $S$ come from the $G^{ST}_1$ and $G^{UT}_1$ terms. Considering the pole at $S \to -T$ of the former and the pole at $S \to 0$ of the latter, and applying \eqref{eq:Integral case 1}, we obtain
\begin{equation}
\label{eq: I2 disk}
    I^{(2)}_{\text{disk}}(T) = T^3 \int_{|z|<1} \d^2z \, \Bigl( |z|^{2T-2}\, |1-z|^{-2T-2}\, G^{ST}_1(z,\bar z) + |1-z|^{-2T-2}\, G^{UT}_1(z,\bar z) \Bigr)\, .
\end{equation}
We note that the disk integral cannot contribute a $\mathcal{O}(T^0)$ term. By virtue of the prefactor $T^3$, we only need to look at the singular terms in $T$ of the integral --- these arise from the behaviour of the integrand at $z = 0,1$. Localising the integration region at $z=0$ and $z=1$, one can check that the weight-six integrands $G^{ST/UT}_1$ degenerate into $\log^6|z|$ and $\zeta(5)\log|z|$ structures around $z = 0$, and $\log^6|1-z|$ and $\zeta(5)\log|1-z|$ structures around $z = 1$. The resulting integrals behave as $T^3 \bigl(\partial_T^6 + \partial_T\bigr) T^{-1} \sim T^{-4} + T$. The singular powers of $T$ can be dropped as part of the supergravity contribution, so the stringy corrections start at the order $\mathcal{O}(T^1)$.

\paragraph{Boundary terms.} The remaining terms of \eqref{eq: G2 channel decomposition} localise on the unit circle via \eqref{eq:boundarygeneral}, with $j$ radial derivatives for each power $S^j$. Absorbing the prefactors $|z|^2$ and $|1-z|^2$ of \eqref{eq: G2 channel decomposition} into the kernels, we have\allowdisplaybreaks{
\begin{align}
\label{eq: I2 circ}
    & I^{(2)}_{\text{circ}}(T) \nonumber \\
    ={}& \oint_{|z|=1} \frac{\d\theta}{2\pi} \biggl[\; T^2\, |1-z|^{-2T-2}\, G^{ST}_2 
    + T \left(\frac r2 \frac{\partial}{\partial r}\right) \frac{|1-z|^{-2T-2}}{|z|^{2}}\, G^{ST}_3 
    + \left(\frac r2 \frac{\partial}{\partial r}\right)^{2} \frac{|1-z|^{-2T-2}}{|z|^{2}}\, G^{ST}_4 \nonumber\\
    & \quad + T^2\, |1-z|^{-2T-2}\, G^{UT}_2 
    + T \left(\frac r2 \frac{\partial}{\partial r}\right) |1-z|^{-2T-2}\, G^{UT}_3 
    + \left(\frac r2 \frac{\partial}{\partial r}\right)^{2} |1-z|^{-2T-2}\, G^{UT}_4 \nonumber\\
    & \quad + \frac1T \left(\frac r2 \frac{\partial}{\partial r}\right)^{3} \frac{|1-z|^{-2T}}{|z|^{2}}\, G^{SU}_1 
    + \left(\frac r2 \frac{\partial}{\partial r}\right)^{2} \frac{|1-z|^{-2T}}{|z|^{2}}\, G^{SU}_2 \nonumber\\
    & \quad + T \left(\frac r2 \frac{\partial}{\partial r}\right) \frac{|1-z|^{-2T}}{|z|^{2}}\, G^{SU}_3 
    + T^2\, |1-z|^{-2T}\, G^{SU}_4 \;\biggr]_{r=1}\, .
\end{align} }
\!\!\!\! By the argument of section~\ref{sec: Second Curvature From Worldsheet}, the worldsheet integrals themselves always have finite $T \to 0$ limits, so only the terms with non-positive powers of $T$ survive: $G^{ST}_4$, $G^{UT}_4$ and $G^{SU}_2$, with prefactor $T^{0}$, and $G^{SU}_1$, with prefactor $1/T$.
After acting with the radial derivatives and restricting to $r=1$ explicitly, we perform the expansion
\begin{equation}
\label{eq: G2 boundary expansions}
    \begin{aligned}
        \left(\frac r2 \frac{\partial}{\partial r}\right)^{2} \frac{|1-z|^{-2T-2}}{|z|^{2}}\, G^{ST}_4 \bigg|_{r=1} &= |1-z|^{-2T-2} \left\lbrack \frac{233 \zeta(3)}{18}-\frac{57913 \zeta(5)}{486} + G^{ST}_{4,\text{reg}} + \mathcal{O}(T) \right\rbrack ,\\
        \left(\frac r2 \frac{\partial}{\partial r}\right)^{2} |1-z|^{-2T-2}\, G^{UT}_4 \bigg|_{r=1} &= |1-z|^{-2T-2} \left\lbrack \frac{233 \zeta(3)}{18}-\frac{57913 \zeta(5)}{486} + G^{UT}_{4,\text{reg}} + \mathcal{O}(T) \right\rbrack ,\\
        \left(\frac r2 \frac{\partial}{\partial r}\right)^{2} \frac{|1-z|^{-2T}}{|z|^{2}}\, G^{SU}_2 \bigg|_{r=1} &= |1-z|^{-2T-2} \left\lbrack \frac{49}{12}\zeta(5) + G^{SU}_{2,\text{reg}} + \mathcal{O}(T) \right\rbrack ,\\
        \frac1T \left(\frac r2 \frac{\partial}{\partial r}\right)^{3} \frac{|1-z|^{-2T}}{|z|^{2}}\, G^{SU}_1 \bigg|_{r=1} &= |1-z|^{-2T-2} \left\lbrack -\frac{233}{12}\zeta(3) + \frac{52621}{324}\zeta(5) + \frac{G^{SU}_{1,\text{sing}}}{T} + G^{SU}_{1,\text{reg}} + \mathcal{O}(T) \right\rbrack ,
    \end{aligned}
\end{equation}
where in each line we have singled out the constant terms as $z= 1$, so that all the functions $G^{\bullet}_{\bullet,\text{reg/sing}}$ generated by the radial derivatives start at order $(z-1)^2$ in the $z \to 1$ limit; we also drop the terms of order $\mathcal{O}(T)$ in the integrand.

The integral of all the constant terms evaluates to
\begin{equation}
\label{eq: I2 constant part}
    \oint_{|z|=1} \frac{\d\theta}{2\pi}\, |1-z|^{-2T-2} \left( \frac{233}{36}\zeta(3) - \frac{17455}{243}\zeta(5) \right) = \frac{\Gamma(-2T-1)}{\Gamma(-T)^2} \left( \frac{233}{9}\zeta(3) - \frac{69820}{243}\zeta(5) \right) ,
\end{equation}
which vanishes as $T \to 0$. For the regular parts $G^{\bullet}_{\bullet,\text{reg}}$ we may set $T=0$ directly, taking $|1-z|^{-2T-2} \to 1/|1-z|^2$, and we find
\begin{equation}
\label{eq: I2 regular parts}
    \begin{aligned}
        \oint_{|z|=1} \frac{\d\theta}{2\pi\, |1-z|^{2}}\, G^{ST}_{4,\text{reg}} &= \frac{\pi^2 \zeta(3)}{12} + \frac{7051 \zeta(3)}{324} - \frac{5\zeta(5)}{6} - \frac{28759 \pi^2}{7776} + \frac{260759 \pi^4}{349920}\,,\\
        \oint_{|z|=1} \frac{\d\theta}{2\pi\, |1-z|^{2}}\, G^{UT}_{4,\text{reg}} &= \frac{\pi^2 \zeta(3)}{12} + \frac{7051 \zeta(3)}{324} - \frac{5\zeta(5)}{6} - \frac{28759 \pi^2}{7776} + \frac{260759 \pi^4}{349920}\,,\\
        \oint_{|z|=1} \frac{\d\theta}{2\pi\, |1-z|^{2}}\, G^{SU}_{2,\text{reg}} &= -\frac{43157 \zeta(3)}{486} - \frac{25999 \pi^4}{38880} + \frac{66227 \pi^2}{5832}\,,\\
        \oint_{|z|=1} \frac{\d\theta}{2\pi\, |1-z|^{2}}\, G^{SU}_{1,\text{reg}} &= \frac{17303 \zeta(3)}{216} - \frac{12215 \pi^4}{23328} - \frac{17465 \pi^2}{5184}\,.
    \end{aligned}
\end{equation}
Finally, the term $\frac1T G^{SU}_{1,\text{sing}}$ in \eqref{eq: G2 boundary expansions} appears singular as $T \to 0$. However, the constant term of its integral vanishes, 
\begin{equation}
    \oint_{|z|=1} \frac{\d\theta}{2\pi\, |1-z|^{2}}\, G^{SU}_{1,\text{sing}} = 0+ \mathcal{O}(T)\,,
\end{equation}
so the full integral remains finite at $T=0$: upon expanding the kernel $|1-z|^{-2T-2} = |1-z|^{-2}\bigl(1 - 2T \log|1-z| + \mathcal{O}(T^2)\bigr)$, the $\mathcal{O}(T)$ term of the integral combines with the $1/T$ prefactor to give the finite contribution 
\begin{equation}
    \frac1T \oint_{|z|=1} \frac{\d\theta}{2\pi\, |1-z|^{2}}\, (-2T) \log(|1-z|)\, G^{SU}_{1,\text{sing}} = -\frac{80347 \zeta(3)}{1944} + \frac{19133 \pi^2}{46656} + \frac{33061 \pi^4}{174960}\,.
\end{equation}
Collecting all terms, we arrive at
\begin{equation}
    I^{(2)}_{\text{circ}}(T) = 6\zeta(2) - \frac{13 \zeta(3)}{2} + \frac{263 \zeta(4)}{6} + \zeta(2)\, \zeta(3) - \frac{5 \zeta(5)}{3} + \mathcal{O}(T^1)\, .
\end{equation}
Since the disk integral starts at order $\mathcal{O}(T)$ as we argued earlier, this constant term of $ I^{(2)}_{\text{circ}}(T)$ fully determines $I^{(2)}(T)\big|_{T^0}$, producing \eqref{eq: I2 T0 from worldsheet} in the main text.

\section{Wilson coefficients of the AdS Virasoro-Shapiro amplitude} \label{app:Wilson-coeff}

Calculating the EEC using the low-energy expansion of Mellin amplitudes requires knowledge of the Wilson coefficients to all orders in this expansion. This appendix is devoted to determining the analytic expressions for the Wilson coefficients.

\subsection{Bootstrap procedure at first curvature}

\label{app: Bootstrap Wilson Coefficients}
We construct an ansatz for the first curvature correction $A^{(1)}(S,T)$ of the scattering of two gravitons with two higher KK modes. The ansatz is constructed using
SVMPLs of transcendental weight 3 with definite parity under $z\leftrightarrow 1-z$. In particular, there are 7 such functions which are constructed explicitly from the weight 3 SVMPL basis in (\ref{eq: SVMPL first curv basis}). Imposing crossing symmetry then gives a linear system for the ansatz coefficients. The remaining constraints follow from matching the poles of the worldsheet expression to the OPE decomposition. Near $S=\delta$, the first curvature correction admits the Laurent
expansion up to a fourth order singularity
\begin{equation}
 A^{(1)}(S,T)
 =
 \sum_{k=1}^{4}
 \frac{\mathcal{R}^{(1)}_k(\delta,T)}
 {(S-\delta)^k}
 +\mathcal{O}\bigl((S-\delta)^0\bigr)\, ,
 \label{eq:A1-pole-expansion}
\end{equation}
where $\mathcal{R}^{(1)}_k$ denotes the physical residue of the
$k$-th order pole. We can match the residues of the worldsheet ansatz to the superblock decomposition which depends on the OPE data\footnote{The twist and reduced OPE coefficient of a heavy single-trace operator \(r=(\delta,\ell,\hat r)\) admit the strong-coupling expansions
\begin{equation*}
    \tau(r;\lambda) = 2\sqrt{\delta}\lambda^{1/4} + \tau_{1}(r) + \tau_{2}(r)\lambda^{-1/4} + \ldots, \quad f(r;\lambda) = f_{0}(r) + f_{1}(r)\lambda^{-1/4} + f_{2}(r)\lambda^{-1/2} + \ldots.
\end{equation*}
Hence \(f_{0}\) is the leading reduced OPE coefficient, \(\tau_{2}\) is the first non-universal curvature correction to the twist, and \(f_{2}\) is the order-\(\lambda^{-1/2}\) correction to the reduced OPE coefficient. Since a single correlator cannot resolve the additional degeneracy label \(\hat r\), the amplitude can only determine the averaged quantities $\langle f_{0}\rangle_{\delta,\ell} = \sum_{\hat r}f_{0}(\delta,\ell,\hat r)$, $\langle f_{0}\tau_{2}\rangle_{\delta,\ell} = \sum_{\hat r}f_{0}(\delta,\ell,\hat r)\tau_{2}(\delta,\ell,\hat r)$ etc.} $\langle f_0\rangle$, $\langle f_0\tau_2\rangle$ and
$\langle f_2\rangle$ to solve these data, see \cite{Alday:2023jdk,Alday:2023mvu} for a detailed discussion.

Following~\cite{Alday:2022xwz, Fardelli:2023fyq} we assume that first curvature Wilson coefficient $\omega^{(1)}_{a,b}(p)$ appeared in \eqref{eq: alt Mellin amp} is in the ring of single-valued multiple zeta values (MZVs) with uniform weight $4+2a+3b$ for all $b \in \mathbb{Z}_{\geq 0}$, $a \in \mathbb{Z}_{\geq -b}$. Using the same conventions as those references, we write the following ansatz 
\begin{align*}
\omega^{(1)}_{a,0}(p)
= 
\sum_{\delta = 1}^\infty \frac{1}{3 \,\delta^{4+a} \, 2^{6+a}}
&\bigl(-16 a^3+2 \bigl(12 a^2+2 a (8 p+33)-8 p^2+16 p-225\bigr) F^{(0)}_1(\delta) \cr & -16 a^2 p-96 a^2+32 a p^2-80 a p-96 a T^{(2)}_0(\delta ) \cr &+16 a+96 F^{(2)}_0(\delta )+32 p^2-64 p-288 T^{(2)}_0(\delta )-831\bigr)
\cr + \,(-1)^{-a} &\bigl(-16 a^3 +2 \bigl(12 a^2+2 a (8 p+33)-8 p^2+16 p-225\bigr) F^{(0)}_1(\delta ) \cr &-16 a^2 p-96 a^2-16 a p^2 +16 a p-96 a T^{(2)}_0(\delta ) \cr &+16 a+96 F^{(2)}_0(\delta )-16 p^2+32 p-288 T^{(2)}_0(\delta )-831\bigr)\, ,
\end{align*}
where we assume the coefficients $F^{(\bullet)}_\bullet(\delta )$ and $T^{(\bullet)}_\bullet(\delta )$ can be written in terms of Euler-Zagier sums~\cite{Alday:2022xwz}
\begin{equation}
\label{eq:Euler-Zagier sums def}
Z_{s_1,\ldots,s_k}(N)
=
\sum_{N \geq n_1 > \cdots > n_k > 0}
\frac{1}{n_1^{s_1}\cdots n_k^{s_k}}\, ,
\qquad
Z(N)=1\, ,
\qquad
Z_{s_1,\ldots,s_k}(0)=0\, .
\end{equation}
We define the depth of $Z_{s_1, \ldots, s_k}(N)$ to be the number of indices, and the weight is given by the sum of indices. When we sum over \(\delta\) we obtain MZVs
\begin{equation} \label{eq:MZVs}
\zeta(s,s_1,s_2,\ldots)
=
\sum_{\delta=1}^{\infty}
\frac{Z_{s_1,s_2,\ldots}(\delta-1)}{\delta^s}\, ,
\end{equation}
thus enforcing our assumption about $\omega^{(1)}_{a,b}(p)$. More concretely,  $F^{(2)}_m(\delta)$ can be written as a linear combination of Euler-Zagier sums with maximum weight $m+3$ and max depth $m+1$, whereas $T^{(2)}_m(\delta)$ has maximum weight $m+2$ and max depth $m+1$. Explicitly we write
\begin{align*}
T^{(2)}_0(\delta)
&=
d_0
+d_1\,\delta\, Z_{1}(\delta-1)
+d_2\,\delta^2\, Z_{2}(\delta-1)\, ,
\\
F^{(2)}_0(\delta)
&=
c_0
+c_1\,\delta\, Z_{1}(\delta-1)
+c_2\,\delta^2\, Z_{2}(\delta-1)
+c_3\,\delta^3\, Z_{3}(\delta-1)
-c_3\,\delta^3\,\zeta(3)\, Z_{0}(\delta-1)\,,
\end{align*}
containing a total of seven unknown coefficients. This is the simplest possible choice such that $\omega^{(1)}_{a,b}(2)$ has enough complexity to match known results from localisation~\cite{Binder:2019jwn, Chester:2020dja}. In particular $\omega^{(1)}_{1,0}(p)$ contains MZVs of depth 2 and hence we must have Euler-Zagier sums of depth 1. All the coefficients $F^{(\bullet)}_\bullet(\delta )$ and $T^{(\bullet)}_\bullet(\delta )$ have definite expressions in terms of the OPE coefficients $\langle f_0\rangle$, $\langle f_0\tau_2\rangle$ and $\langle f_2\rangle$ of $A^{(1)}(S,T)$. The ansatz can then be solved to find
\begin{align}
\label{eq: omega1b=0}
    \omega^{(1)}_{a,0}& = \frac13 \frac{(-1)^a}{2^{2+a}} \Bigl(3 a^2 \zeta(a+3,1)-2 p^2 \zeta(a+3,1) +4 a p \, \zeta(a+3,1) \cr &+(-1)^a \bigl[4 (a+1) p+3 a (a+5)-2 p^2+6\bigr] \zeta(a+3,1) \cr &+4 p \, \zeta(a+3,1)+15 a \, \zeta(a+3,1)-12 ((-1)^a+1) \zeta(a+1,3) \cr &-6 ((-1)^a+1) (a+1) \zeta(a+2,2)+6 \zeta(a+3,1) \cr &+\bigl[-a^3-a^2 p-6 a^2-a p^2 +a p-11 a-p^2+2 p-12\bigr] \zeta (a+4) \cr &+ (-1)^a \bigl[2 (a+1) p^2-(a+1) (a+4) p-a (a (a+6)+11)-12\bigr] \zeta (a+4) \cr &+12 (-1)^a \zeta (3) \zeta (a+1)+12 \zeta (3) \zeta (a+1)\Bigr)\, .
\end{align}
This coefficient should be understood as the sum of an even-in-$a$ and an odd-in-$a$ piece, given explicitly in \eqref{eq: even spin Wilson coeff p22p} and \eqref{eq: odd spin Wilson coeff p22p}.

\subsection{Second curvature results}
\label{app: Bootstrap second curv Wilson coeff}
We can extend the first curvature Wilson coefficient ansatz from the previous subsection to second curvature at $p=2$ as follows {\allowdisplaybreaks
\begin{align}
\alpha^{(2)}_{a,0} = \sum_{\delta=1}^\infty \frac{1}{\delta^{5+2a}} &\Bigg(
F^{(4)}_0(\delta)
-(3+2a) T^{(4)}_0(\delta)
+\frac{1}{4}(3+2a)(7+4a) T^{(2,2)}_0(\delta)\
\\
&+\left(
\frac{141}{16}
+\frac{49a}{12}
+a^2
\right)F^{(2)}_1(\delta)
\cr
&+\left(
\frac{781}{32}
-\frac{a}{3}
-\frac{16a^2}{3}
-\frac{4a^3}{3}
\right)F^{(2)}_0(\delta)
\cr
&+\left(
12
-\frac{167a}{24}
-\frac{73a^2}{6}
-2a^3
\right)T^{(2)}_1(\delta)
\cr
&+\left(
\frac{579}{32}
+\frac{895a}{48}
+22a^2
+16a^3
+\frac{8a^4}{3}
\right)T^{(2)}_0(\delta)
\cr
&+2\left(
-\frac{79693}{1536}
-\frac{553a}{18}
+\frac{613a^2}{288}
+\frac{55a^3}{12}
+\frac{a^4}{2}
\right)F^{(0)}_2(\delta)
\cr
&+\left(
-\frac{425081}{1536}
-\frac{95783a}{1152}
-\frac{461a^2}{96}
-\frac{763a^3}{36}
-\frac{106a^4}{9}
-\frac{4a^5}{3}
\right)F^{(0)}_1(\delta)
\cr
&+\left(
-\frac{1285115}{6144}
+\frac{67307a}{1440}
+\frac{988a^2}{9}
+\frac{495a^3}{8}
+\frac{92a^4}{3}
+\frac{404a^5}{45}
+\frac{8a^6}{9}
\right)
\Bigg)\,, \notag
\end{align} }
\!\!\!\!\! where, as before, all the coefficients $F^{(\bullet)}_\bullet(\delta )$ and $T^{(\bullet)}_\bullet(\delta )$ have explicit expressions in terms of the OPE coefficients of $A^{(2)}(S,T)$ given in~\cite{Alday:2023mvu}. Once again $F^{(\bullet)}_\bullet(\delta )$ and $T^{(\bullet)}_\bullet(\delta )$ are assumed to be expanded in a basis of Euler-Zagier sums. For example,  $T^{(2,2)}_m(\delta)$ has maximum weight $m+6$ and maximum depth $m+2$ such that {\allowdisplaybreaks
\begin{multline}
T^{(2,2)}_0(\delta)= c_0 + c_1 \,\delta\, Z_{1} + \delta^2\!\left(
c_2\, Z_{2}
+c_{11}\, Z_{1,1}
\right)
+\delta^3\!\left(
c_3\, Z_{3}
+c_{21}\, Z_{2,1}
+c_{12}\, Z_{1,2}
\right) \cr
+\delta^4\!\left(
c_4\, Z_{4}
+c_{31}\, Z_{3,1}
+c_{13}\, Z_{1,3}
+c_{22}\, Z_{2,2}
\right)
+\delta^3 \zeta(3)\!\left(
c_{Z3}
+c_{Z31}\,\delta\, Z_{1}
\right) \, , 
\end{multline}
with a total of 13 unknown coefficients. We then solve the bootstrap to find 
\begin{align}
\label{eq: alpha2b0}
\alpha^{(2)}_{a,0} =&\frac{1}{90}
\left(
915 + 4877 a + 8485 a^2 + 6930 a^3 + 3240 a^4 + 808 a^5 + 80 a^6
\right)
\zeta(5+2a)
\cr &+2(4+a)\zeta(2a,5)
+\zeta(-1+2a,6)
+\left(3-\frac{a}{6}-a^2\right)\zeta(1+2a,4)
\cr &+\left(
7-\frac{5a}{6}+2a^2+\frac{2a^3}{3}
\right)\zeta(2+2a,3)
\cr &+\left(
\frac{103}{6}
+\frac{509a}{36}
+\frac{895a^2}{36}
+19a^3
+\frac{11a^4}{3}
\right)\zeta(3+2a,2)
\cr &-\left(
\frac{26}{3}
+\frac{871a}{18}
+\frac{133a^2}{2}
+\frac{560a^3}{9}
+\frac{206a^4}{9}
+\frac{8a^5}{3}
\right)\zeta(4+2a,1)
\cr &+(4+4a)\zeta(2a,2,3)
+(2+8a)\zeta(2a,3,2)
+(-6+12a)\zeta(2a,4,1)
\cr &+4\zeta(-1+2a,3,3)
+6\zeta(-1+2a,4,2)
+12\zeta(-1+2a,5,1)
\cr &+\left(
3-\frac{46a}{3}-4a^2
\right)\zeta(1+2a,1,3)
+\left(
\frac{3}{2}
-\frac{11a}{3}
+2a^2
\right)\zeta(1+2a,2,2)
\cr &-\left(
\frac{3}{2}
+27a
+2a^2
\right)\zeta(1+2a,3,1)
+\left(
\frac{3}{2}
-\frac{14a}{3}
-\frac{52a^2}{3}
-4a^3
\right)\zeta(2+2a,1,2)
\cr &+\left(
2-\frac{35a}{6}-26a^2-6a^3
\right)\zeta(2+2a,2,1)
-4\zeta(-1+2a,3)\zeta(3)
\cr &+\left(
\frac{295}{6}
+\frac{925a}{18}
+\frac{748a^2}{9}
+\frac{104a^3}{3}
+4a^4
\right)\zeta(3+2a,1,1)
\cr &-\left(
6+\frac{8a}{3}+\frac{32a^2}{3}+\frac{8a^3}{3}
\right)\zeta(2+2a)\zeta(3)
-(4+4a)\zeta(2a,2)\zeta(3)
\cr &+\left(
-3+\frac{46a}{3}+4a^2
\right)\zeta(1+2a,1)\zeta(3)
+2\zeta(-1+2a)\zeta(3)^2
-3\zeta(2a)\zeta(5) \, .
\end{align}
}

\section{Localisation of Wilson coefficients and analytic continuation of MZVs}\label{app:ancont}

In section~\ref{sec:low-energy} the strong-coupling coefficients of the EEC are extracted from contour integrals of the Borel resummation of the low-energy expansion. In this appendix we derive the localisation formula quoted in \eqref{eq:localizationresultmain}, together with the regularisation rules required when the resulting individual MZVs are divergent. Consider the general integral
\begin{equation}\label{eq:generalsumint}
    \mathcal{I} \;:=\; \int_{-i\infty}^{+i\infty} \frac{\d s_1}{2\pi i} \sum_{a=0}^{\infty} f(a)\,\frac{\Gamma(2a+p)}{4^a} \zeta(2a+q,\vec{w})\, s_{1}^{2a+m} \,,
\end{equation}
with $\vec{w}=(w_{1},\dots,w_{n})$ a list of positive integers and $p$, $q$, $m$ are also integers. The factor $4^{-a}$ is naturally absorbed by rescaling $u:=s_{1}/2$, so that
\begin{equation}
    \mathcal{I} = 2^{m+1} \int_{-i\infty}^{+i\infty}\!\frac{\d u}{2\pi i}\, \sum_{a=0}^{\infty} f(a)\, \Gamma(2a+p)\,\zeta(2a+q,\vec{w})\, u^{2a+m} \,.
\end{equation}
We use the integral representation of $\zeta(2a+q,\vec{w})$,
\begin{equation}
\label{eq:int rep mzvs}
    \zeta(2a+q,\vec{w}) = \frac{1}{\Gamma(2a{+}q)\,\prod_{j}\Gamma(w_{j})}\int_{0<y_{1}<\dots<y_{n+1}}\!\frac{y_{1}^{2a+q-1}}{e^{y_{1}}-1}\prod_{j=1}^{n}\frac{(y_{j+1}{-}y_{j})^{w_{j}-1}}{e^{y_{j+1}}-1}\,\d y_{1}\cdots\d y_{n+1} \,,
\end{equation}
and interchange the $u$-integral with the $y$-integrals:
\begin{equation}\label{eq:Iintegralgen}
    \begin{aligned}
    \mathcal{I} = \frac{2^{m+1}}{\prod_{j}\Gamma(w_{j})}\int_{0<y_{1}<\dots<y_{n+1}} \! \d^{n+1} y \, \frac{y_{1}^{q-m-1}}{e^{y_{1}}-1}\prod_{j=1}^{n}\frac{(y_{j+1}{-}y_{j})^{w_{j}-1}}{e^{y_{j+1}}-1}  \\
    \times\int_{-i\infty}^{+i\infty}\!\frac{\d u}{2\pi i} \, \sum_{a=0}^{\infty} f(a)\, P(2a)\,(u y_{1})^{2a+m} \,,
    \end{aligned}
\end{equation}
where $P(2a) = \Gamma(2a+p)/\Gamma(2a+q)$. 
The $u$-integral
\begin{equation}
    J(y_{1}) := \int_{-i\infty}^{+i\infty} \frac{\d u}{2\pi i} \, \sum_{a=0}^{\infty} f(a) \, P(2a)\,(u y_{1})^{2a+m}
\end{equation}
can be performed with the following trick. Since $u\,\partial_{u}(uy_{1})^{2a+m}=(2a+m)\,(uy_{1})^{2a+m}$, we may represent the factor $P(2a)$ by the replacement $2a\to u\partial_u-m$.\footnote{When $p \geq q$, $P(2a)$ is a finite polynomial in $a$, and this replacement gives a finite polynomial in the operator $u\partial_u$, as written in \eqref{eq:Popsum}. When $p<q$, $P(2a)$ is rational rather than polynomial, so $P(u\partial_u-m)$ is only an operator notation, involving inverse powers of differential operators. This notation is nevertheless well-defined after contour localisation as indicated in \eqref{eq:Plocalize}: each occurrence of $u\partial_u$ can be replaced by the corresponding numerical factor acting on the residue at $u=1/y_1$. Thus the rational operator is understood through its action on the local residue, rather than as an ordinary differential operator on arbitrary functions.} In particular,
\begin{equation}\label{eq:Popsum}
    \sum_{a=0}^{\infty}f(a)\,P(2a)\,(uy_{1})^{2a+m} = f\left(\frac{u\,\partial_{u}-m}{2}\right) P\left( u\,\partial_{u} - m \right)\,\frac{(u y_1)^{m}}{1-(uy_{1})^{2}}\,,
\end{equation}
so that the $u$-integral becomes
\begin{equation}\label{eq:uintegral}
    J(y_{1}) = \int_{-i\infty}^{+i\infty}\!\frac{\d u}{2\pi i} \, f\left(\frac{u\,\partial_{u}-m}{2}\right)\, P\!\left( u\,\partial_{u} - m \right) \frac{(u y_1)^{m}}{1-(uy_{1})^{2}}  \,.
\end{equation}

\paragraph{Localisation via integration by parts.} Repeated integration by parts in $u$, with vanishing boundary terms at $u\to\pm i\infty$, gives
\begin{equation}
    \int_{-i\infty}^{+i\infty}\!\d u\, (u\partial_{u})^{k} \, g(u) = \bigl(-1\bigr)^{k}\int_{-i\infty}^{+i\infty}\!\d u\, g(u)\, .
\end{equation}
It is important to emphasize that the boundary terms need not vanish for each individual contribution considered here. The integration-by-parts argument should therefore be applied only to the complete Mellin amplitude. Indeed, the full Mellin amplitude is known to decay as $|s|^{-2}$ in the large-$s$ limit ~\cite{Caron-Huot:2021rmr}, even though its individual components may fail to do so. The required cancellation of boundary terms consequently occurs only after all contributions have been combined.

This allows us to deform the contour to the right half plane and localise it on the pole at $u=1/y_1$. On the resulting residue, each occurrence of $u\partial_u$ may be replaced by the numerical factor $-1$. This implies the following substitution rule
\begin{equation}\label{eq:Plocalize}
    f\!\left(\frac{u\,\partial_{u}-m}{2}\right) P\!\left( u\,\partial_{u} - m \right) \;\longrightarrow\; f\!\left(-\frac{m+1}{2}\right) P\!\left(-1-m\right) \,.
\end{equation}
Plugging \eqref{eq:Plocalize} and \eqref{eq:uintegral} back into \eqref{eq:Iintegralgen}, we obtain:
\begin{equation}\label{eq:localizationresult}
    \begin{aligned}
        \mathcal{I} \;&=\; \frac{2^{m}}{\prod_{j}\Gamma(w_{j})}\; f\!\left(-\frac{m+1}{2}\right) P\!\left(-m-1\right)\; \int_{0<y_{1}<\dots<y_{n+1}} \frac{y_{1}^{\,q-m-2}}{e^{y_{1}}-1}\prod_{j=1}^{n}\frac{(y_{j+1}{-}y_{j})^{w_{j}-1}}{e^{y_{j+1}}-1}\,\d^{n+1} y \\
        &=\; 2^{m}\, f\!\left(-\frac{m+1}{2}\right)\Gamma(p-m-1)\, \zeta(q-m-1,\,\vec{w})\,.
    \end{aligned}
\end{equation}
Equation (\ref{eq:localizationresult}) is the localisation statement: the infinite sum over $a$ collapses to a single MZV evaluated at the shifted leading argument $q-m-1$. However, when $q-m-1 \leq 1$, the MZV $\zeta(q-m-1,\vec{w})$ is divergent and must be regulated.

\paragraph{Divergent leading entry.} To make sense of $\zeta(q-m-1,\vec{w})$ with $q-m-1\leq 1$, we perform regularisation via the shift $m \to m-2\varepsilon$, so that the reduced MZV has a Laurent expansion in $\varepsilon$:
\begin{equation}\label{eq:Iepsilon}
    \mathcal{I}_{\varepsilon} = 2^{m-2\varepsilon}\, f\!\left(-\frac{m+1}{2}+\varepsilon\right) \Gamma(p-m-1+2\varepsilon)\,\zeta(q-m-1+2\varepsilon,\,\vec{w})\,.
\end{equation}
For $q-m-1 \leq 0$, we use the same integral representation as in the first line of \eqref{eq:localizationresult}, now with $n_{1}:=q-m-1\leq 0$. The factor $y_{1}^{2\varepsilon}$ regulates the lower endpoint of the innermost integral and turns the would-be divergence into an explicit $1/\varepsilon$ pole. Peeling off the $y_{1}$-integral, binomial-expanding $(y_{2}-y_{1})^{w_{1}-1}$, and applying the Bernoulli expansion $1/(e^{y}-1)=\sum_{k\geq 0}\frac{B_{k}}{k!}y^{k-1}$, we perform the $y_1$-integral term by term, and the remaining $y$-integrals reassemble into a depth-lowered MZV:
\begin{equation}\label{eq:negativeentryrecursion}
    \begin{aligned}
        \zeta(n_{1}+2\varepsilon,\,\vec{w})
        = \sum_{k=0}^{k_{\max}} & \frac{B_{k}}{k!} \, \frac{\Gamma(n_{1}{+}w_{1} {+} 2\varepsilon {+} k {-} 1)}{\Gamma(n_{1}+2\varepsilon)\,\Gamma(w_{1})} \left\lbrack \sum_{i=0}^{w_{1}-1}\binom{w_{1} {-} 1}{i}\frac{(-1)^{i}}{n_{1}{+}2\varepsilon {+} i {+} k {-} 1} \right\rbrack \\ 
        & \times \zeta(n_{1}+w_{1}+2\varepsilon+k-1,\,w_{2},\dots,w_{n}) \,, \quad n_{1} \leq 0 \,.
    \end{aligned}
\end{equation}
Since we only care about the limit $\varepsilon\to 0$, the summation over $k$ is effectively truncated: the prefactor $1/\Gamma(n_{1}+2\varepsilon)$ vanishes at order $\varepsilon^{1}$, so only the terms singular in $\varepsilon$ survive, and these are capped at 
\begin{equation}
    k_{\max}=\max(1-n_{1},\,2-n_{1}-w_{1}) \,.
\end{equation}
This provides a recursion relation in terms of MZVs with the depth lowered by one, so that the recursion terminates.

Iterating \eqref{eq:negativeentryrecursion} gives convergent MZVs, together with divergent MZVs of the form $\zeta(1+2\varepsilon,\dots)$; the same objects arise directly from \eqref{eq:Iepsilon} in the boundary case $q-m-1=1$. They can be evaluated via the stuffle relation and represented by convergent MZVs and $\zeta(1+2\varepsilon,1,\dots,1)$:
\begin{equation}\label{eq:stufflereg}
\zeta\bigl(\underbrace{1+2\varepsilon,1,\dots,1}_{\ell},\,\mathbf{b}\bigr)
    = \sum_{i=0}^{\ell}\zeta\bigl(\underbrace{1+2\varepsilon,1,\dots,1}_{i}\bigr)\,
    \zeta_{\text{reg}}\bigl(\underbrace{1,\dots,1}_{\ell-i},\,\mathbf{b}\bigr) \,, \quad b_{1}\geq 2\,,
\end{equation}
with the convention $\zeta()\equiv 1$.

The divergent factor $\zeta(1+2\varepsilon,1,\dots,1)$ has the Laurent expansion
\begin{equation}\label{eq:zeta1laurent}
\zeta\bigl(1+2\varepsilon,\underbrace{1,\dots,1}_{\ell}\bigr) = \frac{1}{\Gamma(1+2\varepsilon)\,(2\varepsilon)^{1+\ell}} + \mathcal{O}(\varepsilon)\,.
\end{equation}
In the end, $\mathcal{I}_{\varepsilon}$ has a well-defined Laurent expansion around $\varepsilon = 0$. The individual terms may carry poles in $\varepsilon$, but after summing all the contributions to a given Wilson coefficient $\alpha^{(k)}_{a,b}$, the singular terms cancel, leaving a finite and well-defined result. The final result is a linear combination of ordinary MZVs.

This completes the evaluation of \eqref{eq:generalsumint}: the localisation formula \eqref{eq:localizationresult}, together with the regularisation rules \eqref{eq:negativeentryrecursion}, \eqref{eq:stufflereg} and \eqref{eq:zeta1laurent}, suffices to evaluate all the low-energy integrals encountered. The localisation point is $a=-\frac{m+1}{2}$: for $m=0$ this reproduces the $a=-\frac12$ localisation used for $c_{0,0}$, $c_{1,1}$ and $c_{2,2}$ in sections~\ref{sec: Analytic specialization} and~\ref{sec: Second Curvature Low Energy}, and for $m=2$ the $a=-\frac32$ localisation used for $c_{2,1}$ in section~\ref{sec: Second Curvature Low Energy}.

\bibliography{refs}

@article{Ren:2026zxs,
    author = "Ren, Lecheng and Wang, Bo and Wen, Congkao",
    title = "{Energy-energy correlator from the AdS Virasoro-Shapiro amplitude}",
    eprint = "2601.05312",
    archivePrefix = "arXiv",
    primaryClass = "hep-th",
    doi = "10.1103/m9q6-47z2",
    journal = "Phys. Rev. D",
    volume = "113",
    number = "10",
    pages = "106013",
    year = "2026"
}

@article{Alday:2023jdk,
    author = "Alday, Luis F. and Hansen, Tobias and Silva, Joao A.",
    title = "{Emergent Worldsheet for the AdS Virasoro-Shapiro Amplitude}",
    eprint = "2305.03593",
    archivePrefix = "arXiv",
    primaryClass = "hep-th",
    doi = "10.1103/PhysRevLett.131.161603",
    journal = "Phys. Rev. Lett.",
    volume = "131",
    number = "16",
    pages = "161603",
    year = "2023"
}

@article{Chester:2020vyz,
    author = "Chester, Shai M. and Green, Michael B. and Pufu, Silviu S. and Wang, Yifan and Wen, Congkao",
    title = "{New modular invariants in $ \mathcal{N} $ = 4 Super-Yang-Mills theory}",
    eprint = "2008.02713",
    archivePrefix = "arXiv",
    primaryClass = "hep-th",
    reportNumber = "PUPT-2620, QMUL-PH-20-16",
    doi = "10.1007/JHEP04(2021)212",
    journal = "JHEP",
    volume = "04",
    pages = "212",
    year = "2021"
}

@article{Brown:2023why,
    author = "Brown, Augustus and Wen, Congkao and Xie, Haitian",
    title = "{Generating functions and large-charge expansion of integrated correlators in {\ensuremath{\mathscr{N}}} = 4 supersymmetric Yang-Mills theory}",
    eprint = "2303.17570",
    archivePrefix = "arXiv",
    primaryClass = "hep-th",
    reportNumber = "QMUL-PH-23-03",
    doi = "10.1007/JHEP07(2023)129",
    journal = "JHEP",
    volume = "07",
    pages = "129",
    year = "2023"
}

@article{Caetano:2023zwe,
    author = "Caetano, Jo{\~a}o and Komatsu, Shota and Wang, Yifan",
    title = "{Large charge {\textquoteright}t Hooft limit of $ \mathcal{N} $ = 4 super-Yang-Mills}",
    eprint = "2306.00929",
    archivePrefix = "arXiv",
    primaryClass = "hep-th",
    reportNumber = "CERN-TH-2023-089",
    doi = "10.1007/JHEP02(2024)047",
    journal = "JHEP",
    volume = "02",
    pages = "047",
    year = "2024"
}

@article{Jiang:2019xdz,
    author = "Jiang, Yunfeng and Komatsu, Shota and Vescovi, Edoardo",
    title = "{Structure constants in $ \mathcal{N} $ = 4 SYM at finite coupling as worldsheet g-function}",
    eprint = "1906.07733",
    archivePrefix = "arXiv",
    primaryClass = "hep-th",
    reportNumber = "CERN-TH-2019-093, Imperial-TP-EV-2019-01",
    doi = "10.1007/JHEP07(2020)037",
    journal = "JHEP",
    volume = "07",
    number = "07",
    pages = "037",
    year = "2020"
}

@article{Chen:2025yxg,
    author = "Chen, Junding and Jiang, Yunfeng and Zhou, Xinan",
    title = "{Giant Graviton Correlators as Defect Systems}",
    eprint = "2503.22987",
    archivePrefix = "arXiv",
    primaryClass = "hep-th",
    reportNumber = "USTC-ICTS/PCFT-25-14",
    doi = "10.1103/hg9p-hblr",
    journal = "Phys. Rev. Lett.",
    volume = "135",
    number = "8",
    pages = "081602",
    year = "2025"
}

@article{Brown:2025cbz,
    author = "Brown, Augustus and Galvagno, Francesco and Grassi, Alba and Iossa, Cristoforo and Wen, Congkao",
    title = "{Large charge meets semiclassics in $\mathcal{N}$ = 4 super Yang-Mills}",
    eprint = "2503.02028",
    archivePrefix = "arXiv",
    primaryClass = "hep-th",
    reportNumber = "CERN-TH-2025-016, QMUL-PH-25-03",
    doi = "10.1007/JHEP06(2025)223",
    journal = "JHEP",
    volume = "06",
    pages = "223",
    year = "2025"
}

@article{Chester:2024wnb,
    author = "Chester, Shai M. and Zhong, De-liang",
    title = "{AdS3{\texttimes}S3 Virasoro-Shapiro Amplitude with Ramond-Ramond Flux}",
    eprint = "2412.06429",
    archivePrefix = "arXiv",
    primaryClass = "hep-th",
    doi = "10.1103/PhysRevLett.134.151602",
    journal = "Phys. Rev. Lett.",
    volume = "134",
    number = "15",
    pages = "151602",
    year = "2025"
}

@article{Jiang:2025oar,
    author = "Jiang, Hongliang and Zhong, De-liang",
    title = "{AdS$_{3}${\texttimes} S$^{3}$ Virasoro-Shapiro amplitude with KK modes}",
    eprint = "2508.06039",
    archivePrefix = "arXiv",
    primaryClass = "hep-th",
    doi = "10.1007/JHEP02(2026)071",
    journal = "JHEP",
    volume = "02",
    pages = "071",
    year = "2026"
}

@article{Virasoro:1969me,
    author = "Virasoro, M. A.",
    title = "{Alternative constructions of crossing-symmetric amplitudes with regge behavior}",
    doi = "10.1103/PhysRev.177.2309",
    journal = "Phys. Rev.",
    volume = "177",
    pages = "2309--2311",
    year = "1969"
}

@article{Schlotterer:2012ny,
    author = "Schlotterer, O. and Stieberger, S.",
    title = "{Motivic Multiple Zeta Values and Superstring Amplitudes}",
    eprint = "1205.1516",
    archivePrefix = "arXiv",
    primaryClass = "hep-th",
    reportNumber = "AEI-2012-039, MPP-2012-859",
    doi = "10.1088/1751-8113/46/47/475401",
    journal = "J. Phys. A",
    volume = "46",
    pages = "475401",
    year = "2013"
}

@article{Broedel:2013tta,
    author = "Broedel, Johannes and Schlotterer, Oliver and Stieberger, Stephan",
    title = "{Polylogarithms, Multiple Zeta Values and Superstring Amplitudes}",
    eprint = "1304.7267",
    archivePrefix = "arXiv",
    primaryClass = "hep-th",
    reportNumber = "DAMTP-2013-22, AEI-2013-194, MPP-2013-119",
    doi = "10.1002/prop.201300019",
    journal = "Fortsch. Phys.",
    volume = "61",
    pages = "812--870",
    year = "2013"
}

@article{Chester:2025ssu,
    author = "Chester, Shai M. and Ferrero, Pietro and Pavarini, Daniele R.",
    title = "{Modular invariant gluon-graviton scattering in AdS at one loop}",
    eprint = "2504.10319",
    archivePrefix = "arXiv",
    primaryClass = "hep-th",
    doi = "10.1007/JHEP08(2025)208",
    journal = "JHEP",
    volume = "08",
    pages = "208",
    year = "2025"
}

@article{Caron-Huot:2021rmr,
    author = "Caron-Huot, Simon and Mazac, Dalimil and Rastelli, Leonardo and Simmons-Duffin, David",
    title = "{Sharp boundaries for the swampland}",
    eprint = "2102.08951",
    archivePrefix = "arXiv",
    primaryClass = "hep-th",
    doi = "10.1007/JHEP07(2021)110",
    journal = "JHEP",
    volume = "07",
    pages = "110",
    year = "2021"
}

@article{Cuomo:2025pjp,
    author = "Cuomo, Gabriel and Firat, Eren and Nardi, Filippo and Ricci, Lorenzo",
    title = "{Conformal collider physics at large charge}",
    eprint = "2503.21867",
    archivePrefix = "arXiv",
    primaryClass = "hep-th",
    doi = "10.1007/JHEP02(2026)223",
    journal = "JHEP",
    volume = "02",
    pages = "223",
    year = "2026"
}

@article{DeLillo:2025stg,
    author = "De Lillo, L. and Duan, Z. and Frau, M. and Galvagno, F. and Lerda, A. and Vallarino, P. and Wen, C.",
    title = "{$ \mathcal{N}=2 $ universality at strong coupling}",
    eprint = "2510.27594",
    archivePrefix = "arXiv",
    primaryClass = "hep-th",
    reportNumber = "QMUL-PH-25-32",
    doi = "10.1007/JHEP02(2026)019",
    journal = "JHEP",
    volume = "02",
    pages = "019",
    year = "2026"
}

@article{Brown:2013gia,
    author = "Brown, Francis",
    title = "{Single-valued Motivic Periods and Multiple Zeta Values}",
    eprint = "1309.5309",
    archivePrefix = "arXiv",
    primaryClass = "math.NT",
    doi = "10.1017/fms.2014.18",
    journal = "SIGMA",
    volume = "2",
    pages = "e25",
    year = "2014"
}

@article{Chester:2020dja,
    author = "Chester, Shai M. and Pufu, Silviu S.",
    title = "{Far beyond the planar limit in strongly-coupled $ \mathcal{N} $ = 4 SYM}",
    eprint = "2003.08412",
    archivePrefix = "arXiv",
    primaryClass = "hep-th",
    reportNumber = "PUPT-2616",
    doi = "10.1007/JHEP01(2021)103",
    journal = "JHEP",
    volume = "01",
    pages = "103",
    year = "2021"
}

@article{Shapiro:1970gy,
    author = "Shapiro, Joel A.",
    title = "{Electrostatic analog for the virasoro model}",
    reportNumber = "MDDP-TR-71-039",
    doi = "10.1016/0370-2693(70)90255-8",
    journal = "Phys. Lett. B",
    volume = "33",
    pages = "361--362",
    year = "1970"
}

@article{Dixon:2019uzg,
    author = "Dixon, Lance J. and Moult, Ian and Zhu, Hua Xing",
    title = "{Collinear limit of the energy-energy correlator}",
    eprint = "1905.01310",
    archivePrefix = "arXiv",
    primaryClass = "hep-ph",
    reportNumber = "SLAC-PUB-17427, SLAC--PUB--17427",
    doi = "10.1103/PhysRevD.100.014009",
    journal = "Phys. Rev. D",
    volume = "100",
    number = "1",
    pages = "014009",
    year = "2019"
}

@article{Chester:2024esn,
    author = "Chester, Shai M. and Hansen, Tobias and Zhong, De-liang",
    title = "{The type IIA Virasoro-Shapiro amplitude in AdS$_{4}$ {\texttimes} CP$^{3}$ from ABJM theory}",
    eprint = "2412.08689",
    archivePrefix = "arXiv",
    primaryClass = "hep-th",
    doi = "10.1007/JHEP05(2025)040",
    journal = "JHEP",
    volume = "05",
    pages = "040",
    year = "2025"
}

@article{Brown:2024yvt,
    author = "Brown, Augustus and Galvagno, Francesco and Wen, Congkao",
    title = "{All-loop Heavy-Heavy-Light-Light correlators in $ \mathcal{N} $ = 4 super Yang-Mills theory}",
    eprint = "2407.02250",
    archivePrefix = "arXiv",
    primaryClass = "hep-th",
    reportNumber = "QMUL-PH-24-12",
    doi = "10.1007/JHEP10(2024)171",
    journal = "JHEP",
    volume = "10",
    pages = "171",
    year = "2024"
}

@article{Jiang:2023uut,
    author = "Jiang, Yunfeng and Wu, Yu and Zhang, Yang",
    title = "{Giant correlators at quantum level}",
    eprint = "2311.16791",
    archivePrefix = "arXiv",
    primaryClass = "hep-th",
    reportNumber = "USTC-ICTS/PCFT-23-35",
    doi = "10.1007/JHEP05(2024)345",
    journal = "JHEP",
    volume = "05",
    pages = "345",
    year = "2024"
}

@article{He:2026ios,
    author = "He, Song and Shi, Canxin and Tang, Yichao and Wen, Congkao",
    title = "{Bootstrapping Giant Graviton Correlators}",
    eprint = "2605.14281",
    archivePrefix = "arXiv",
    primaryClass = "hep-th",
    reportNumber = "QMUL-PH-26-17",
    month = "5",
    year = "2026"
}

@article{Corley:2001zk,
    author = "Corley, Steve and Jevicki, Antal and Ramgoolam, Sanjaye",
    title = "{Exact correlators of giant gravitons from dual N=4 SYM theory}",
    eprint = "hep-th/0111222",
    archivePrefix = "arXiv",
    reportNumber = "BROWN-HET-1292",
    doi = "10.4310/ATMP.2001.v5.n4.a6",
    journal = "Adv. Theor. Math. Phys.",
    volume = "5",
    pages = "809--839",
    year = "2002"
}

@article{Hashimoto:2000zp,
    author = "Hashimoto, Akikazu and Hirano, Shinji and Itzhaki, N.",
    title = "{Large branes in AdS and their field theory dual}",
    eprint = "hep-th/0008016",
    archivePrefix = "arXiv",
    reportNumber = "NSF-ITP-00-063",
    doi = "10.1088/1126-6708/2000/08/051",
    journal = "JHEP",
    volume = "08",
    pages = "051",
    year = "2000"
}

@article{McGreevy:2000cw,
    author = "McGreevy, John and Susskind, Leonard and Toumbas, Nicolaos",
    title = "{Invasion of the giant gravitons from Anti-de Sitter space}",
    eprint = "hep-th/0003075",
    archivePrefix = "arXiv",
    reportNumber = "SU-ITP-00-09",
    doi = "10.1088/1126-6708/2000/06/008",
    journal = "JHEP",
    volume = "06",
    pages = "008",
    year = "2000"
}

@article{Paul:2023rka,
    author = "Paul, Hynek and Perlmutter, Eric and Raj, Himanshu",
    title = "{Exact large charge in $ \mathcal{N} $ = 4 SYM and semiclassical string theory}",
    eprint = "2303.13207",
    archivePrefix = "arXiv",
    primaryClass = "hep-th",
    doi = "10.1007/JHEP08(2023)078",
    journal = "JHEP",
    volume = "08",
    pages = "078",
    year = "2023"
}

@article{Aprile:2024lwy,
    author = "Aprile, Francesco and Giusto, Stefano and Russo, Rodolfo",
    title = "{Holographic correlators with BPS bound states in $\mathcal{N} = 4$ SYM}",
    eprint = "2409.12911",
    archivePrefix = "arXiv",
    primaryClass = "hep-th",
    doi = "10.1103/PhysRevLett.134.091602",
    journal = "Phys. Rev. Lett.",
    volume = "134",
    number = "9",
    pages = "091602",
    year = "2025"
}

@article{Chester:2019jas,
    author = "Chester, Shai M. and Green, Michael B. and Pufu, Silviu S. and Wang, Yifan and Wen, Congkao",
    title = "{Modular invariance in superstring theory from $ \mathcal{N} $ = 4 super-Yang-Mills}",
    eprint = "1912.13365",
    archivePrefix = "arXiv",
    primaryClass = "hep-th",
    reportNumber = "PUPT-2607, QMUL-PH-19-37",
    doi = "10.1007/JHEP11(2020)016",
    journal = "JHEP",
    volume = "11",
    pages = "016",
    year = "2020"
}

@article{Montonen:1977sn,
    author = "Montonen, C. and Olive, David I.",
    title = "{Magnetic Monopoles as Gauge Particles?}",
    reportNumber = "CERN-TH-2391",
    doi = "10.1016/0370-2693(77)90076-4",
    journal = "Phys. Lett. B",
    volume = "72",
    pages = "117--120",
    year = "1977"
}

@article{Brown:2024tru,
    author = "Brown, Augustus and Galvagno, Francesco and Wen, Congkao",
    title = "{Exact results for giant graviton four-point correlators}",
    eprint = "2403.17263",
    archivePrefix = "arXiv",
    primaryClass = "hep-th",
    doi = "10.1007/JHEP07(2024)049",
    journal = "JHEP",
    volume = "07",
    pages = "049",
    year = "2024"
}

@article{Brown:2026dhy,
    author = "Brown, Augustus and Dorigoni, Daniele and Wen, Congkao",
    title = "{Giant graviton integrated correlators at finite coupling and all orders in $1/N$}",
    eprint = "2603.20083",
    archivePrefix = "arXiv",
    primaryClass = "hep-th",
    reportNumber = "QMUL-PH-26-11",
    month = "3",
    year = "2026"
}

@article{Hashimoto:1996bf,
    author = "Hashimoto, Akikazu and Klebanov, Igor R.",
    editor = "Dijkgraaf, R. and Klebanov, Igor R. and Narain, K. S. and Randjbar-Daemi, S.",
    title = "{Scattering of strings from D-branes}",
    eprint = "hep-th/9611214",
    archivePrefix = "arXiv",
    reportNumber = "PUPT-1669",
    doi = "10.1016/S0920-5632(97)00074-1",
    journal = "Nucl. Phys. B Proc. Suppl.",
    volume = "55",
    pages = "118--133",
    year = "1997"
}

@article{Goncalves:2014ffa,
    author = "Gon{\c{c}}alves, Vasco",
    title = "{Four point function of $\mathcal{N}=4$ stress-tensor multiplet at strong coupling}",
    eprint = "1411.1675",
    archivePrefix = "arXiv",
    primaryClass = "hep-th",
    doi = "10.1007/JHEP04(2015)150",
    journal = "JHEP",
    volume = "04",
    pages = "150",
    year = "2015"
}

@article{Belitsky:2013xxa,
    author = "Belitsky, A. V. and Hohenegger, S. and Korchemsky, G. P. and Sokatchev, E. and Zhiboedov, A.",
    title = "{From correlation functions to event shapes}",
    eprint = "1309.0769",
    archivePrefix = "arXiv",
    primaryClass = "hep-th",
    reportNumber = "CERN-PH-TH-2013-211, IPHT-T13-210, LAPTH-047-13",
    doi = "10.1016/j.nuclphysb.2014.04.020",
    journal = "Nucl. Phys. B",
    volume = "884",
    pages = "305--343",
    year = "2014"
}

@article{Mack:2009mi,
    author = "Mack, Gerhard",
    title = "{D-independent representation of Conformal Field Theories in D dimensions via transformation to auxiliary Dual Resonance Models. Scalar amplitudes}",
    eprint = "0907.2407",
    archivePrefix = "arXiv",
    primaryClass = "hep-th",
    month = "7",
    year = "2009"
}

@article{Penedones:2010ue,
    author = "Penedones, Joao",
    title = "{Writing CFT correlation functions as AdS scattering amplitudes}",
    eprint = "1011.1485",
    archivePrefix = "arXiv",
    primaryClass = "hep-th",
    doi = "10.1007/JHEP03(2011)025",
    journal = "JHEP",
    volume = "03",
    pages = "025",
    year = "2011"
}

@article{Kologlu:2019mfz,
    author = "Kologlu, Murat and Kravchuk, Petr and Simmons-Duffin, David and Zhiboedov, Alexander",
    title = "{The light-ray OPE and conformal colliders}",
    eprint = "1905.01311",
    archivePrefix = "arXiv",
    primaryClass = "hep-th",
    reportNumber = "CALT-TH 2019-013, CERN-TH-2019-055",
    doi = "10.1007/JHEP01(2021)128",
    journal = "JHEP",
    volume = "01",
    pages = "128",
    year = "2021"
}

@article{Basham:1977iq,
    author = "Basham, C. Louis and Brown, Lowell S. and Ellis, S. D. and Love, S. T.",
    title = "{Electron - Positron Annihilation Energy Pattern in Quantum Chromodynamics: Asymptotically Free Perturbation Theory}",
    reportNumber = "RLO-1388-746",
    doi = "10.1103/PhysRevD.17.2298",
    journal = "Phys. Rev. D",
    volume = "17",
    pages = "2298",
    year = "1978"
}

@article{Basham:1978bw,
    author = "Basham, C. Louis and Brown, Lowell S. and Ellis, Stephen D. and Love, Sherwin T.",
    title = "{Energy Correlations in electron - Positron Annihilation: Testing QCD}",
    reportNumber = "RLO-1388-759",
    doi = "10.1103/PhysRevLett.41.1585",
    journal = "Phys. Rev. Lett.",
    volume = "41",
    pages = "1585",
    year = "1978"
}

@article{Sterman:1975xv,
    author = "Sterman, George F.",
    title = "{Jet Structure in e+ e- Annihilation with Massless Hadrons}",
    reportNumber = "ILL-TH-75-32",
    month = "12",
    year = "1975"
}

@article{Dempsey:2025yiv,
    author = "Dempsey, Ross and Karlsson, Robin and Pufu, Silviu S. and Zahraee, Zahra and Zhiboedov, Alexander",
    title = "{Conformal collider bootstrap in $\mathcal{N}=4$ SYM}",
    eprint = "2512.10796",
    archivePrefix = "arXiv",
    primaryClass = "hep-th",
    year = "2025"
}

@article{Fardelli:2023fyq,
    author = "Fardelli, Giulia and Hansen, Tobias and Silva, Joao A.",
    title = "{AdS Virasoro-Shapiro amplitude with KK modes}",
    eprint = "2308.03683",
    archivePrefix = "arXiv",
    primaryClass = "hep-th",
    doi = "10.1007/JHEP11(2023)064",
    journal = "JHEP",
    volume = "11",
    pages = "064",
    year = "2023"
}

@article{Alday:2023mvu,
    author = "Alday, Luis F. and Hansen, Tobias",
    title = "{The AdS Virasoro-Shapiro amplitude}",
    eprint = "2306.12786",
    archivePrefix = "arXiv",
    primaryClass = "hep-th",
    doi = "10.1007/JHEP10(2023)023",
    journal = "JHEP",
    volume = "10",
    pages = "023",
    year = "2023"
}

@article{Hofman:2008ar,
    author = "Hofman, Diego M. and Maldacena, Juan",
    title = "{Conformal collider physics: Energy and charge correlations}",
    eprint = "0803.1467",
    archivePrefix = "arXiv",
    primaryClass = "hep-th",
    doi = "10.1088/1126-6708/2008/05/012",
    journal = "JHEP",
    volume = "05",
    pages = "012",
    year = "2008"
}

@article{Eden:2000bk,
    author = "Eden, Burkhard and Petkou, Anastasios C. and Schubert, Christian and Sokatchev, Emery",
    title = "{Partial nonrenormalization of the stress tensor four point function in N=4 SYM and AdS / CFT}",
    eprint = "hep-th/0009106",
    archivePrefix = "arXiv",
    reportNumber = "LAPTH-811-2000, KL-TH-00-06",
    doi = "10.1016/S0550-3213(01)00151-1",
    journal = "Nucl. Phys. B",
    volume = "607",
    pages = "191--212",
    year = "2001"
}

@article{Nirschl:2004pa,
    author = "Nirschl, M. and Osborn, H.",
    title = "{Superconformal Ward identities and their solution}",
    eprint = "hep-th/0407060",
    archivePrefix = "arXiv",
    reportNumber = "DAMTP-04-51",
    doi = "10.1016/j.nuclphysb.2005.01.013",
    journal = "Nucl. Phys. B",
    volume = "711",
    pages = "409--479",
    year = "2005"
}

@article{Chicherin:2023gxt,
    author = "Chicherin, Dmitry and Korchemsky, Gregory P. and Sokatchev, Emery and Zhiboedov, Alexander",
    title = "{Energy correlations in heavy states}",
    eprint = "2306.14330",
    archivePrefix = "arXiv",
    primaryClass = "hep-th",
    reportNumber = "CERN-TH-2023-109, IPhT-T23/051, LAPTH-034/23",
    doi = "10.1007/JHEP11(2023)134",
    journal = "JHEP",
    volume = "11",
    pages = "134",
    year = "2023"
}

@article{Belitsky:2013bja,
    author = "Belitsky, A. V. and Hohenegger, S. and Korchemsky, G. P. and Sokatchev, E. and Zhiboedov, A.",
    title = "{Event shapes in $\mathcal{N} = 4$ super-Yang-Mills theory}",
    eprint = "1309.1424",
    archivePrefix = "arXiv",
    primaryClass = "hep-th",
    reportNumber = "CERN-PH-TH-2013-212",
    doi = "10.1016/j.nuclphysb.2014.04.019",
    journal = "Nucl. Phys. B",
    volume = "884",
    pages = "206--256",
    year = "2014"
}

@article{Dorigoni:2021guq,
    author = "Dorigoni, Daniele and Green, Michael B. and Wen, Congkao",
    title = "{Exact properties of an integrated correlator in $ \mathcal{N} $ = 4 SU(N) SYM}",
    eprint = "2102.09537",
    archivePrefix = "arXiv",
    primaryClass = "hep-th",
    reportNumber = "QMUL-PH-21-09, DCPT-21/03",
    doi = "10.1007/JHEP05(2021)089",
    journal = "JHEP",
    volume = "05",
    pages = "089",
    year = "2021"
}

@article{Brown_2004,
     author = {Francis C.S. Brown},
     title = {Polylogarithmes multiples uniformes en une variable},
     journal = {Comptes Rendus. Math\'ematique},
     pages = {527--532},
     year = {2004},
     publisher = {Elsevier},
     volume = {338},
     number = {7},
     doi = {10.1016/j.crma.2004.02.001},
     language = {fr},
}

@article{Brown_2020,
      title={Depth-graded motivic multiple zeta values}, 
      author={Francis Brown},
      year={2020},
      eprint={1301.3053v2},
      archivePrefix={arXiv},
      primaryClass={math.NT},
      url={https://arxiv.org/abs/1301.3053}, 
}

@article{Wang:2025pjo,
    author = "Wang, Bo and Wu, Di and Yuan, Ellis Ye",
    title = "{Kaluza-Klein AdS Virasoro-Shapiro Amplitude near Flat Space}",
    eprint = "2503.01964",
    archivePrefix = "arXiv",
    primaryClass = "hep-th",
    doi = "10.1103/v72s-rv7y",
    journal = "Phys. Rev. Lett.",
    volume = "135",
    number = "4",
    pages = "041603",
    year = "2025"
}

@article{Collins:1981uk,
    author = "Collins, John C. and Soper, Davison E.",
    title = "{Back-To-Back Jets in QCD}",
    reportNumber = "OITS-155",
    doi = "10.1016/0550-3213(81)90339-4",
    journal = "Nucl. Phys. B",
    volume = "193",
    pages = "381",
    year = "1981",
    note = "[Erratum: Nucl.Phys.B 213, 545 (1983)]"
}

@article{Korchemsky:2019nzm,
    author = "Korchemsky, G. P.",
    title = "{Energy correlations in the end-point region}",
    eprint = "1905.01444",
    archivePrefix = "arXiv",
    primaryClass = "hep-th",
    reportNumber = "IPhT-T19/041",
    doi = "10.1007/JHEP01(2020)008",
    journal = "JHEP",
    volume = "01",
    pages = "008",
    year = "2020"
}

@article{Alday:2022xwz,
    author = "Alday, Luis F. and Hansen, Tobias and Silva, Joao A.",
    title = "{AdS Virasoro-Shapiro from single-valued periods}",
    eprint = "2209.06223",
    archivePrefix = "arXiv",
    primaryClass = "hep-th",
    doi = "10.1007/JHEP12(2022)010",
    journal = "JHEP",
    volume = "12",
    pages = "010",
    year = "2022"
}

@article{Duhr:2019tlz,
    author = "Duhr, Claude and Dulat, Falko",
    title = "{PolyLogTools {\textemdash} polylogs for the masses}",
    eprint = "1904.07279",
    archivePrefix = "arXiv",
    primaryClass = "hep-th",
    reportNumber = "CP3-19-17, CERN-TH-2019-045, SLAC-PUB-17423",
    doi = "10.1007/JHEP08(2019)135",
    journal = "JHEP",
    volume = "08",
    pages = "135",
    year = "2019"
}

@article{Belitsky:2014zha,
    author = "Belitsky, A. V. and Hohenegger, S. and Korchemsky, G. P. and Sokatchev, E.",
    title = "{N=4 superconformal Ward identities for correlation functions}",
    eprint = "1409.2502",
    archivePrefix = "arXiv",
    primaryClass = "hep-th",
    reportNumber = "CERN-PH-TH-2014-174, IPHT-T14-122, LAPTH-109-14",
    doi = "10.1016/j.nuclphysb.2016.01.008",
    journal = "Nucl. Phys. B",
    volume = "904",
    pages = "176--215",
    year = "2016"
}

@article{Binder:2019jwn,
    author = "Binder, Damon J. and Chester, Shai M. and Pufu, Silviu S. and Wang, Yifan",
    title = "{$ \mathcal{N} $ = 4 Super-Yang-Mills correlators at strong coupling from string theory and localization}",
    eprint = "1902.06263",
    archivePrefix = "arXiv",
    primaryClass = "hep-th",
    reportNumber = "PUPT-2582",
    doi = "10.1007/JHEP12(2019)119",
    journal = "JHEP",
    volume = "12",
    pages = "119",
    year = "2019"
}

@article{Chen:2023wah,
    author = "Chen, Hao and Zhou, Xinan and Zhu, Hua Xing",
    title = "{Power corrections to energy flow correlations from large spin perturbation}",
    eprint = "2301.03616",
    archivePrefix = "arXiv",
    primaryClass = "hep-ph",
    doi = "10.1007/JHEP10(2023)132",
    journal = "JHEP",
    volume = "10",
    pages = "132",
    year = "2023"
}

@article{Chen:2026lic,
    author = "Chen, Hao and Moult, Ian and Wang, Bo and Zhu, Hua Xing",
    title = "{The Back-to-Back Plateau of the Energy-Energy Correlator from Weak to Strong Coupling}",
    eprint = "2610.07145",
    archivePrefix = "arXiv",
    primaryClass = "hep-th",
    journal = "",
    reportNumber = "MIT-CTP 6124",
    month = "10",
    year = "2026"
}

@article{Fa:2026xxx,
    author = "Fa, Xiaoyu and Wang, Bo and Yuan, Ellis Ye",
    eprint = "26xx.xxxxx",
    archivePrefix = "arXiv",
    primaryClass = "hep-ph",
    title = "{World-sheet Bootstrap of Kaluza-Klein AdS Virasoro-Shapiro Amplitude, to appear}",
    journal = "",
}

@article{Basham:1978zq,
    author = "Basham, C. L. and Brown, L. S. and Ellis, S. D. and Love, S. T.",
    title = "{Energy Correlations in electron-Positron Annihilation in Quantum Chromodynamics: Asymptotically Free Perturbation Theory}",
    reportNumber = "RLO-1388-761",
    doi = "10.1103/PhysRevD.19.2018",
    journal = "Phys. Rev. D",
    volume = "19",
    pages = "2018",
    year = "1979"
}

@article{Chen:2020vvp,
    author = "Chen, Hao and Moult, Ian and Zhang, XiaoYuan and Zhu, Hua Xing",
    title = "{Rethinking jets with energy correlators: Tracks, resummation, and analytic continuation}",
    eprint = "2004.11381",
    archivePrefix = "arXiv",
    primaryClass = "hep-ph",
    doi = "10.1103/PhysRevD.102.054012",
    journal = "Phys. Rev. D",
    volume = "102",
    number = "5",
    pages = "054012",
    year = "2020"
}

@article{Komiske:2022enw,
    author = "Komiske, Patrick T. and Moult, Ian and Thaler, Jesse and Zhu, Hua Xing",
    title = "{Analyzing N-Point Energy Correlators inside Jets with CMS Open Data}",
    eprint = "2201.07800",
    archivePrefix = "arXiv",
    primaryClass = "hep-ph",
    reportNumber = "MIT-CTP 5389",
    doi = "10.1103/PhysRevLett.130.051901",
    journal = "Phys. Rev. Lett.",
    volume = "130",
    number = "5",
    pages = "051901",
    year = "2023"
}

@article{Moult:2025nhu,
    author = "Moult, Ian and Zhu, Hua Xing",
    title = "{Energy Correlators: A Journey From Theory to Experiment}",
    eprint = "2506.09119",
    archivePrefix = "arXiv",
    primaryClass = "hep-ph",
    journal = "",
    month = "6",
    year = "2025"
}

@article{Belitsky:2013ofa,
    author = "Belitsky, A. V. and Hohenegger, S. and Korchemsky, G. P. and Sokatchev, E. and Zhiboedov, A.",
    title = "{Energy-Energy Correlations in N=4 Supersymmetric Yang-Mills Theory}",
    eprint = "1311.6800",
    archivePrefix = "arXiv",
    primaryClass = "hep-th",
    reportNumber = "CERN-PH-TH-2013-282, IPHT-13-264, LAPTH-069-13",
    doi = "10.1103/PhysRevLett.112.071601",
    journal = "Phys. Rev. Lett.",
    volume = "112",
    number = "7",
    pages = "071601",
    year = "2014"
}

@article{Henn:2019gkr,
    author = "Henn, J. M. and Sokatchev, E. and Yan, K. and Zhiboedov, A.",
    title = "{Energy-energy correlation in $N$=4 super Yang-Mills theory at next-to-next-to-leading order}",
    eprint = "1903.05314",
    archivePrefix = "arXiv",
    primaryClass = "hep-th",
    reportNumber = "CERN-TH-2019-026, LAPTH-014/19, MPP-2019-54",
    doi = "10.1103/PhysRevD.100.036010",
    journal = "Phys. Rev. D",
    volume = "100",
    number = "3",
    pages = "036010",
    year = "2019"
}

@article{Chang:2025kgq,
    author = "Chang, Cyuan-Han and Chen, Hao and Liu, Xiaohui and Simmons-Duffin, David and Yuan, Feng and Zhu, Hua Xing",
    title = "{Quantum Scaling in Energy Correlators beyond the Confinement Transition}",
    eprint = "2507.15923",
    archivePrefix = "arXiv",
    primaryClass = "hep-ph",
    reportNumber = "MIT-CTP 5894, CALT-TH 2025-023, CPTNP-2025-025",
    doi = "10.1103/9ml8-xkfc",
    journal = "Phys. Rev. Lett.",
    volume = "136",
    number = "8",
    pages = "081903",
    year = "2026"
}

@article{Lee:2025okn,
    author = "Lee, Kyle and Stewart, Iain W.",
    title = "{Dihadron Fragmentation and the Confinement Transition in Energy Correlators}",
    eprint = "2507.11495",
    archivePrefix = "arXiv",
    primaryClass = "hep-ph",
    reportNumber = "MIT-CTP 5889",
    doi = "10.1103/m18j-xypt",
    journal = "Phys. Rev. Lett.",
    volume = "136",
    number = "8",
    pages = "081902",
    year = "2026"
}

@article{Guo:2025zwb,
    author = "Guo, Yuxun and Yuan, Feng and Zhao, Wenbin",
    title = "{Factorization and Resummation for the Nearside Energy-Energy Correlators}",
    eprint = "2507.15820",
    archivePrefix = "arXiv",
    primaryClass = "hep-ph",
    doi = "10.1103/4qkq-x5st",
    journal = "Phys. Rev. Lett.",
    volume = "136",
    number = "8",
    pages = "081904",
    year = "2026"
}

@article{Kang:2025zto,
    author = "Kang, Zhong-Bo and Metz, Andreas and Pitonyak, Daniel and Zhang, Congyue",
    title = "{Dihadron Fragmentation Framework for Near-Side Energy-Energy Correlators}",
    eprint = "2507.17444",
    archivePrefix = "arXiv",
    primaryClass = "hep-ph",
    doi = "10.1103/jnl4-x77t",
    journal = "Phys. Rev. Lett.",
    volume = "136",
    number = "8",
    pages = "081905",
    year = "2026"
}

@article{Jiang:2026xnh,
    author = "Jiang, Hongliang",
    title = "{D1-D5 CFT data from AdS$_{3}${\texttimes} S$^{3}$ Virasoro-Shapiro amplitude}",
    eprint = "2601.18646",
    archivePrefix = "arXiv",
    primaryClass = "hep-th",
    doi = "10.1007/JHEP06(2026)077",
    journal = "JHEP",
    volume = "06",
    pages = "077",
    year = "2026"
}

@article{Wu:2025ott,
    author = "Wu, Yu and Jiang, Yunfeng and Liu, Chang and Zhang, Yang",
    title = "{All giant graviton two-point functions at two-loops}",
    eprint = "2509.23161",
    archivePrefix = "arXiv",
    primaryClass = "hep-th",
    reportNumber = "USTC-ICTS/PCFT-25-35",
    doi = "10.1007/JHEP03(2026)097",
    journal = "JHEP",
    volume = "03",
    pages = "097",
    year = "2026"
}

\end{document}